\documentstyle[preprint,aps]{revtex}  

\newcommand\Det{{\rm Det}}

\newcommand\tr{{\rm tr}}

\newcommand\Tr{{\rm Tr}}

\begin{document}

\draft

\title{Spin Factor in Path Integral Representation for Dirac
Propagator in External  Fields}

\author{Dmitri M. Gitman}

\address{Instituto de F\'{\i}sica, Universidade de S\~ao Paulo \\ 
Caixa Postal 66318-CEP 05389-970-S\~ao Paulo, S.P., Brazil \\
}

\author{Stoian I. Zlatev\footnote{On leave from the Institute for
Nuclear Research and Nuclear Energy, Sofia, Bulgaria}}

\address{Departamento de F\'{\i}sica,
Universidade Federal de Sergipe\\
49000-000 Aracaju, SE, Brazil \\}

\date{\today}

\maketitle

\begin{abstract}
We study the   spin factor problem both in $3+1$ and $2+1$
dimensions  which are essentially different for spin factor construction. 
Doing all Grassmann integrations in the corresponding  path integral 
representations for Dirac propagator we get representations with  spin factor 
in arbitrary external field. Thus, the propagator appears to be
presented by means of  bosonic path integral only. In $3+1$ dimensions  we present  a simple
derivation of  spin factor avoiding some unnecessary steps 
in the original brief letter (Gitman, Shvartsman, Phys. Lett. {\bf B318}
(1993) 122) which themselves need some additional justification.
In this way the meaning of the surprising possibility
of complete integration over Grassmann variables gets clear. 
In $2+1$ dimensions the derivation of the spin factor is completely
original. Then we use the representations with spin factor for
calculations of the propagator in some configurations of external
fields. Namely, in  constant uniform electromagnetic
field and in its combination  with a plane wave field.  
\end{abstract}

\newpage

\section{Introduction}

Propagators of relativistic particles in external fields 
(electromagnetic, non-Abelian or 
gravitational) contain important information about the quantum behavior of 
these particles. 
Moreover, if such propagators are 
known in arbitrary external field one can find exact 
one-particle Green functions in the corresponding quantum field theory 
taking functional integrals over all  external fields. Dirac propagator 
in an external electromagnetic field distinguishes from that of a scalar 
particle by a complicated spinor structure. The problem of its path
integral representation has attracted researchers' attention  
already for a long time.  Thus, Feynman who has written first his path
integral for the probability amplitude in nonrelativistic quantum mechanics
\cite{F1} and then wrote a 
path integral for the causal Green function of Klein-Gordon
equation (scalar particle propagator) 
\cite{F2}, had also made an attempt to derive a representation for Dirac 
propagator via a bosonic path integral \cite{F3}. 
After the introduction of the integral over Grassmann variables by 
Berezin it turned to be possible to present this propagator via both bosonic
and Grassmann  variables, the latter describe
spinning degrees of freedom. Representations of this kind have been
discussed in the literature for a long time in different contexts 
\cite{all1}. Nevertheless, attempts to
write Dirac  propagator via only a bosonic path integral continued.
 Thus, Polyakov \cite{b12} assumed that the propagator 
of a free Dirac electron in $D=3$ Euclidean space-time can 
be presented by means of a bosonic path integral similar to the 
scalar particle case, modified by 
so called spin factor (SF). This idea was developed in \cite{b13}
e.g. to write SF for Dirac fermions, interacting with a non-Abelian gauge 
field in $D$-dimensional Euclidean space-time. In those representations  
SF itself was presented via some additional bosonic path
integrals and its $\gamma$-matrix structure was not defined
explicitly. Surprisingly, it was shown in  \cite{b20} that all 
Grassmann integrations in the representation of Dirac propagator
in an arbitrary external field  in $3+1$ dimensions can be done, so that an 
expression for SF was derived as a
given functional of the bosonic trajectory. Having such representation
with SF, one can use it to calculate the propagator in some
particular cases of external fields. This way of calculation
provides automatically
an explicit spinor structure of the propagators which can be used for
concrete calculations in the Furry picture (see for example, 
\cite{b9,Odin}).  

In the recent work \cite{Git} the propagator of a spinning particle in
an external field was presented via a path integral in arbitrary
dimensions. It turns out that the problem has different solutions in
even and odd dimensions. In even dimensions the representation is just
a generalization of one in four dimensions mentioned above. In odd
dimensions the solution was presented for the first time and differs
essentially from the even-dimensional case. Using the representation
in odd dimensions one can derive an expression for SF doing
Grassmann integrations similar to the four-dimensional case \cite{b20}.  

In the present paper we continue the consideration of the problems related
to the SF conception. Namely, we discuss derivation of SF
both in even and odd dimensions on the examples of $3+1$ and
$2+1$ cases and then we use the path integral representations with
SF to calculate the propagators in some configurations of
external fields. In $3+1$ dimensions  we present  a simple
derivation of SF  avoiding some unnecessary steps 
in the original brief letter \cite{b20} which 
themselves needed some additional justification.
In this way the meaning of the surprising possibility
of complete integration over Grassmann variables gets clear. 
Then we use the representation with SF for
calculations of the propagator in   a constant uniform electromagnetic
field and its combination with a plane wave. Due to the fact that this way
of calculations  provides automatically  an
explicit $\gamma$-matrix structure of the propagator,  the representations
obtained   differ from
those found by means of other methods, for example differs from the
well known Schwinger formula in the constant uniform electromagnetic
field. To compare both representations we prove in the Appendix some complicated
decompositions of functions on the $\gamma$ matrices. In $2+1$
dimensions the derivation of SF is completely original. We
calculate then the propagator in these dimensions in constant
electromagnetic field by means of the representation with SF. 
The result is new and cannot be derived from the $3+1$-dimensional 
case by means of a dimensional reduction.

\section{Spin factor in $3+1$ dimensions}

\subsection{Doing integrals over Grassmann variables}

The propagator of a relativistic spinning particle in an external electromagnetic
field $A_{\mu}(x)$ is the causal Green function $S^{c}(x,y)$ of the Dirac 
equation in this field,
\begin{equation}
\label{e1}
\left[\gamma^{\mu}\left(i\partial_{\mu}-gA_{\mu}\right)-m\right]S^{c}(x,y)
=-\delta^{4}(x-y)\;,\;
\end{equation}
where $x=(x^{\mu}),\;\left[\gamma^{\mu}\;,\;\gamma^{\nu}\right]_{+}
=2\eta^{\mu \nu};\;\eta^{\mu \nu}={\rm diag}(1,-1,-1,-1);\;
\mu,\nu=0,1,2,3$.

In the paper \cite{b10} the following  Lagrangian path integral 
representation  for the propagator was obtained in $3+1$ dimensions, 
\begin{eqnarray}\label{e2}
&&S^{c}=S^{c}(x_{out},x_{in})=-\tilde{S}^{c}\gamma^5\;, \\
&&\tilde{S}^{c}=\exp\left\{i\tilde{\gamma}^{n}
\frac{\partial_{\ell}}{\partial \theta^{n}}\right\}\int_{0}^{\infty}de_{0}
\int d\chi_{0}\int_{e_{0}}
M(e)De \int_{\chi_{0}}D\chi \int_{x_{in}}^{x_{out}}Dx \int
D\pi_{e}\int D\pi_{\chi} \nonumber \\
&&\times \int_{\psi(0)+\psi(1)=\theta} 
{\cal D}\psi  
\exp\left\{i\int_{0}^{1}\left[-\frac{\dot{x}^{2}}
{2e}-\frac{e}{2}m^{2}-g\dot{x}^{\mu}A_{\mu}+iegF_{\mu \nu}\psi^{\mu}\psi^{\nu}\right.
\right.\nonumber \\
&&\left.\left.\left. +i\left(\frac{\dot{x}_{\mu}\psi^{\mu}}{e}-m\psi^{5}\right)
\chi-i\psi_{n}\dot{\psi}^{n}+\pi_{e} \dot{e}+\pi_{\chi} \dot{\chi}\right]d\tau
+ \psi_{n}(1)\psi^{n}(0)\right\}\right|_{\theta=0}\;, \nonumber
\end{eqnarray}
\noindent where $\tilde{\gamma}^{\mu}=\gamma^5
\gamma^{\mu},\;\tilde{\gamma}^{5}=\tilde{\gamma}^0 \tilde{\gamma}^1 \tilde{\gamma}^2
\tilde{\gamma}^3=
\gamma^0 \gamma^1 \gamma^2 \gamma^3=\gamma^5,\;\;\left[\gamma^{m}\;,\;
\gamma^{n}\right]_{+}=2\eta^{mn},
\; m,n=0,1,2,3,5, \; \; \eta^{mn}={\rm diag}(1,-1,-1,-1,-1)$;
\noindent $\theta^{n}$ are auxiliary Grassmann (odd) variables, 
anticommuting by definition with the $\gamma$-matrices; $x^{\mu}(\tau), 
\; e(\tau), \; \pi_{e}(\tau)$ are bosonic trajectories of 
integration; $\psi^{n}(\tau), \; \chi(\tau), \; \pi_{\chi}(\tau)$ are odd 
trajectories of integration; boundary conditions
\[x(0)=x_{in}, \;x(1)=x_{out},\;
e(0)=e_{0},\;\psi^{n}(0)+\psi^{n}(1)=\theta^{n}, \; 
\chi(0)=\chi_{0} 
\]
take place; the measure $M(e)$ and ${\cal D}\psi$
have the form
\begin{equation}\label{e3}
M(e)=\int Dp\exp\left\{ \frac{i}{2}\int_{0}^{1} ep^{2}d\tau\right\}, \;
{\cal D}\psi=D\psi\left[\int_{\psi(0)+\psi(1)=0} D\psi \exp\left\{\int_{0}^{1}
 \psi_{n}\dot{\psi}^{n}
d\tau \right\}\right]^{-1},
\end{equation}
and $\frac{\partial_{\ell}}{\partial\theta^{n}}$ stands for the left 
derivatives.
Let us demonstrate that the  propagator (\ref{e2}) can 
be expressed only through a bosonic path integral over the
coordinates $x$. 
For this purpose one needs to perform several functional 
integrations, in particular, to fulfil all the Grassmann
integrations. First,
one can integrate over $\pi_{e}$ and $\pi_{\chi}$, 
and then using the arising $\delta$-functions to remove the functional 
integration over $e$ and 
$\chi$, 
\begin{eqnarray*}
&&\tilde{S}^{c}=-\exp\left\{i\tilde{\gamma}^{n}
\frac{\partial_{\ell}}{\partial \theta^{n}}\right\}
\int_{0}^{\infty}de_{0}\;M(e_{0})\int_{x_{in}}^{x_{out}}Dx\; 
\int_{\psi(0)+\psi(1)=\theta} {\cal D}\psi
\;\int_{0}^{1} \left(\frac{\dot{x}_{\mu}\psi^{\mu}}{e_{0}}-
m\psi^{5}\right)d\tau \\ 
&&\left.\times \exp\left\{i\int_{0}^{1}\left[-\frac{\dot{x}^{2}}{2e_{0}}-
\frac{e_{0}}{2}m^{2}-g\dot{x}^{\mu}A_{\mu}+ige_{0}F_{\mu \nu}\psi^{\mu}\psi^{\nu}-
i\psi_{n}\dot{\psi}^{n}\right]d\tau+ \psi_{n}(1)\psi^{n}(0)\right\}
\right|_{\theta=0}.
\end{eqnarray*}
Then, changing the integration variables,
\begin{equation}\label{e4}
\psi\sp{n}= \frac{1}{2}\left(\xi\sp{n}+\theta\sp{n}\right),
\end{equation}
and introducing odd sources $\rho\sb{n}(\tau)$ for the new variables $\xi\sp{n}(\tau)$,
we get
\begin{eqnarray}\label{e5}
&&\tilde{S}^{c}=-\frac{1}{2}\exp\left\{i\tilde{\gamma}^{n}
\frac{\partial_{\ell}}{\partial \theta^{n}}\right\}
\int_{0}^{\infty}de_{0}\; M(e_{0})\int_{x_{in}}^{x_{out}}Dx\; 
\exp\left\{-i\left[\frac{\dot{x}_{\mu}\star \dot{x}^{\mu}}{2e_{0}}+\frac{e_{0}}{2}
m^{2}+g\dot{x}^{\mu}\star A_{\mu}\right]\right.  \nonumber \\
&&-\left.\left.\frac{ge\sb{0}}{4}\theta\sp{\mu}\star{\cal
F}\sb{\mu\nu}\star\theta\sp{\nu}\right\}
\left[\frac{\dot{x}_{\mu}}{e_{0}}\star\left(\frac{\delta_{\ell}}
{\delta \rho_{\mu}}+\theta^{\mu}\right)
-m\star\left(\frac{\delta_{\ell}}{\delta \rho_{5}}+\theta^{5}\right)  
\right]R[x,\rho,\theta]\right\vert_{\rho=0,\theta=0},
\end{eqnarray}  
\noindent where  
\begin{eqnarray}
\label{e5a}
&&R[x,\rho,\theta]=\int_{\xi(0)+\xi(1)=0}{\cal
D}\xi\,\exp\left\{\frac{1}{4}
\xi_{n}\star\dot\xi^n\right.     \nonumber                  \\
&&-\left.\frac{ge_{0}}{4}\xi^{\mu}\star{\cal F}_{\mu\nu}
\star\xi^{\nu}-\frac{ge_{0}}{2}\theta^{\mu}\star{\cal F}_{\mu\nu}
\star\xi^{\nu}+\rho_{n}\star\xi^{n}\right\}\,,         \\
&&{\cal D}\xi=D\xi\left[\int_{\xi(0)+\xi(1)=0}D\xi\,
\exp\left\{\frac{1}{4} \xi_{n}\star
\dot\xi^{n}\right\}\right]^{-1}\;. 
\end{eqnarray}
Here  condensed notations are used in which ${\cal F}_{\mu\nu}$ is 
understood as a matrix with continuous indices,
\begin{equation}
\label{e5b}
{\cal F}_{\mu\nu}(\tau,\tau^\prime)=F_{\mu\nu}(x(\tau))
\delta(\tau-\tau^\prime)\,,
\end{equation}
and integration over $\tau$ is denoted by star, e.g.
\[
\xi_{n}\star\dot\xi^{n}=\int_{0}^{1}\xi_{n}(\tau)\dot\xi^{n}(\tau)d\tau \,.
\]
Sometimes discrete indices will be also omitted. In this case all
tensors of second rank have to be understood as matrices with lines
marked by the  first contravariant indices of the tensors, and with
columns marked by the  second covariant  indices  of the tensors. 

The Grassmann  Gaussian path integral in (\ref{e5a}) can be evaluated
straightforwardly \cite{b19} to be 
\begin{equation}
\label{e6}
R[x,\rho,\theta]=\left\{\Det \left[{\cal U}^{-1}(0){\cal U}(g)
\right]\right\}^{1/2}\exp\left\{J^{m}\star {\cal W}_{mn}\star 
J^{n}\right\}\,,
\end{equation}  
where the matrices  ${\cal W}(g)$ and $
{\cal U}(g)$ have the form
\begin{eqnarray}
&&{\cal W}_{mn}(g)=\left(\begin{array}{cc}
{\cal U}_{\mu\nu}(g)  &        0                                     \\
0                 &        -\delta^\prime(\tau-\tau^\prime)
\end{array}\right)\,,\nonumber\\
\label{e5c}
&&{\cal U}_{\mu\nu}(g)=\eta_{\mu\nu}\delta^\prime(\tau-\tau^\prime)
-ge_{0}{\cal F}_{\mu\nu}(\tau, \tau^\prime)\,,
\end{eqnarray}
and 
\[ 
J_{\mu}=\rho_{\mu}+\frac{ge_{0}}{2}{\cal F}_{\mu\nu}\star
\theta^{\nu},\qquad  J_{5}=\rho_{5}.
\]
The determinant in (\ref{e6}) should be  understood as 
\begin{eqnarray}
&&\Det\left[{\cal U}^{-1}(0){\cal U}(g)\right]=\exp{\rm Tr}\left[\log{\cal U}
(g)-\log{\cal U}(0)\right]  \nonumber  \\ 
\label{e7}
&&=\exp \left\{-e_{0}{\rm Tr}\int_{0}^{g}dg^{\prime}{\cal R}
(g^{\prime})\star{\cal F}\right\}\,,
\end{eqnarray}
where ${\cal R}(g)$ is the inverse to  ${\cal U}(g)$,
considered as an operator acting in the space of the antiperiodic
functions,
\begin{eqnarray}
&&\frac{d}{d\tau}{\cal R}_{\mu\nu}(g\vert\tau,\tau^{\prime})-ge_{0}
F_{\mu}^{\quad\lambda}(x(\tau)){\cal R}_{\lambda\nu}(g\vert\tau,\tau^\prime)
=\eta_{\mu\nu}\delta(\tau-\tau^\prime),\nonumber\\
\label{e8}
&&{\cal R}_{\mu\nu}(g\vert 1,\tau)=-{\cal R}_{\mu\nu}(g\vert 0,\tau)\,, 
\qquad \forall\tau\in (0,\,1).
\end{eqnarray}
Substituting   (\ref{e6}) and (\ref{e7}) into (\ref{e5}) and
performing then the 
functional differentiations with respect to $\rho_n$, we get
\begin{eqnarray}\label{e9}
&&\tilde{S}^c=-\frac{1}{2}\exp\left\{i\tilde{\gamma}^{n}\frac{\partial
_{\ell}}{\partial\theta^{n}}\right\}\int_{0}^{\infty}de_{0}M(e_{0})
\int_{x_{in}}^{x_{out}}Dx\exp\left\{-\frac{i}{2}\left[
\frac{\dot{x}^{\mu}\star\dot{x}^{\mu}}{e_{0}}+e_{0}m^{2}\right.\right.\nonumber  \\
&&+\left.\left.g\dot{x}^{\mu}
A_{\mu}\right]\right\}              
\left[\frac{\dot{x}^{\mu}}{e_{0}}\star K_{\mu\nu}\theta^{\nu}
-m\theta^{5}\right]\left[1-\frac{ge_{0}}{4}B_{\alpha\beta}\theta^{\alpha}
\theta^{\beta}+\frac{g^{2}e_{0}^{2}}{16}B_{\alpha\beta}B^{*\alpha\beta}
\theta^{0}\theta^{1}\theta^{2}\theta^{3}\right] \nonumber            \\
&&\times\left.\exp\left\{-\frac{e_0}{2}\int_{0}^{\infty}dg^{\prime}
{\rm Tr} {\cal R}(g^{\prime})\star{\cal F}\right\}\right\vert_{\theta=0}\,, 
\end{eqnarray}
where following notations are used,
\begin{equation}
\label{e9a}
B_{\mu\nu}=F_{\mu\lambda}\star K^{\lambda}_{\quad\nu},\quad
B^{*\mu\nu}=\frac{1}{2}\epsilon^{\alpha\beta\mu\nu}B_{\alpha\beta},
\quad K_{\mu\nu}=\eta_{\mu\nu}+ge_{0}{\cal R}_{\mu\lambda}(g)
\star F^{\lambda}_{\quad\nu}\,,
\end{equation}
and $\epsilon^{\mu\nu\alpha\beta}$ is Levi-Civita symbol normalized by
$\epsilon^{0123}=1$.  

Differentiation with respect to $\theta^{n}$ in (\ref{e9}) replaces
the products of the  variables $\theta^{n}$ by the corresponding
antisymmetrized products of the matrices $i\tilde{\gamma}^{n}$. Finally,
passing to the propagator $S^{c}$ and using the identities
\begin{equation}
\label{e9b}
\gamma^{[\lambda}\gamma^{\mu}\gamma^{\nu]}=
\gamma^{\lambda}\gamma^{[\mu}\gamma^{\nu]}
-2\eta^{\lambda[\mu}\gamma^{\nu]},
\qquad \sigma^{\mu\nu}=i\gamma^{[\mu}\gamma^{\nu]},\nonumber
\end{equation}
where antisymmetrization over the corresponding sets of indices is
denoted by brackets, one gets
\begin{equation}
\label{e10}
S^{c}(x_{out},x_{in})=\frac{i}{2}\int_{0}^{\infty}de_{0}\int_
{x_{in}}^{x_{out}}Dx\,M(e_{0})\Phi[x,e_{0}]\,\exp\left\{iI[x,e_{0}]
\right\}\,,
\end{equation}
where $I[x,e_{0}]$ is the action of a relativistic spinless particle,
\begin{equation}
\label{e11}
I[x,e_{0}]=-\int_{0}^{1}\left[\frac{\dot{x}^{2}}{2e_{0}}+\frac{e_{0}}
{2}m^{2}+g\dot{x}A(x)\right]\, d\tau\,,
\end{equation} 
and $\Phi[x,e_{0}]$ is SF,
\begin{eqnarray}
\label{e13}
&&\Phi\lbrack x,\, e\sb0\rbrack=\left\lbrack  
m+ (2e\sb0)\sp{-1}\dot x\sp\mu
\star K\sb{\mu\lambda}\left(2\eta\sp{\lambda\kappa}-ge\sb0
B\sp{\lambda\kappa}\right)\gamma\sb{\kappa}\right.\nonumber\\
&&\left.-\frac{{\rm i}g}{4}\left(
me\sb0+\dot x\sp\mu\star K\sb{\mu\lambda}\gamma\sp\lambda\right) 
B\sb{\kappa\nu}\sigma\sp{\kappa\nu}
+m\frac{g\sp2e\sb0\sp2}{16} B\sb{\alpha\beta}\sp* B\sp{\alpha\beta}
\gamma\sp5\right\rbrack 
\exp\left\{-\frac{e\sb0}{2}\int\sb0\sp g dg\sp\prime {\rm Tr\,}
Q(g\sp\prime)\star{\cal  F}
\right\}\,.      
\end{eqnarray}

\subsection{Propagator in constant uniform electromagnetic field}

In the case of a constant uniform field ($F\sb{\mu\nu}=const$), which we 
are going to discuss in this section, the functionals  ${\cal R}$, $K$ and
$B$ do not depend on the
trajectory $x$ and can be calculated 
straightforwardly,
\begin{eqnarray}
\label{e17}
&&{\cal R}(g)=
\frac{1}{2}\left(\eta\varepsilon(\tau-\tau\sp\prime)
-\tanh\frac{ge\sb0 F}{2}\right) 
\exp\{e\sb0 gF(\tau-\tau\sp\prime)\}\,,  \nonumber    \\
&&K=\left(\eta-\tanh\frac{ge\sb0 F}{2}\right)
\exp\left(ge\sb0 F\tau\right)\,, \qquad B=\frac{2}{ge\sb0}
\tanh\frac{ge\sb0 F}{2}\,.                
\end{eqnarray}
Using them in (\ref{e13}) and integrating over $\tau$ whenever 
possible, we obtain SF in the constant uniform field,
\begin{eqnarray}\label{e18}
&&\Phi\lbrack x,\,e\sb0\rbrack=\left({\rm det}\cosh\frac{ge\sb0 F}{2}\right) 
\sp{1/2}\left\{m\left\lbrack 1-\frac{ i}{2}\left(\tanh\frac{ge\sb0
F}{2}\right)\sb{\mu\nu}\sigma\sp{\mu\nu}\right.\right.         \nonumber \\
&&+\left.\frac{1}{4}
\left(\tanh\frac {ge\sb0 F}{2}\right)\sb{\mu\nu}\sp\ast
\left(\tanh\frac {ge\sb0 F}{2}\right)\sp{\mu\nu}\gamma\sp5\right\rbrack 
+\frac{1}{e\sb0}\left(\int\sb0\sp1\dot x\exp
(ge\sb0 F\tau)d\tau\right)\nonumber\\
&&\left.\times\left(\eta-\tanh\frac{ge\sb0F}{2}\right)
\left\lbrack\left(\eta-\tanh\frac{ge\sb0F}{2}\right)\gamma
-\frac{i}{2}
\gamma\left(\tanh\frac{ge\sb0F}{2}\right)\sb{\mu\nu}\sigma
\sp{\mu\nu}\right\rbrack\right\}\,.           
\end{eqnarray} 
We can see that in the field under consideration SF is
linear in the trajectory $x\sp\mu(\tau)$. That facilitates the bosonic
integration in the expression (\ref{e10}).

In spite of the fact that SF is a gauge invariant object, 
the total
propagator is not. It is clear from the expression (\ref{e10}) where one
needs to choose a particular gauge for the potentials
$A\sb\mu$. Namely, we are going to use the following potentials
\begin{equation}
\label{e19}
A\sb\mu=-\frac{1}{2}F\sb{\mu\nu}x\sp\nu\,,    
\end{equation}
for the constant uniform field $F\sb{\mu\nu}=const$. 
Thus, one can see that the path integral (\ref{e10}) is quasi-Gaussian
in the case under consideration.  
Let us make there the shift $x\rightarrow y+x\sb{cl}$,
with $x\sb{cl}$ a solution of the classical equations of motion 
\begin{equation}\label{e20}
\frac{\delta I}{\delta x}=0
\quad\Leftrightarrow\quad
\ddot x\sb{\mu}-ge\sb{0}F\sb{\mu\nu}\dot x\sp{\nu} =0\,
\end{equation}
subjected to the boundary conditions
$x\sb{cl}(0)=x\sb{in}\,, \quad x\sb{cl}(1)=x\sb{out}$.
Then the new trajectories of integration $y$ obey zero boundary
conditions, $y(0)=y(1)=0$.
Due to the quadratic structure of the action $I\lbrack x,e\sb0
\rbrack$ and the linearity of SF  in $x$ one
can make the following substitutions in the path integral:
\begin{eqnarray}
\label{e21}
&&I\lbrack y+x\sb{cl},\,e\sb0\rbrack\rightarrow 
I\lbrack x\sb{cl},\,e\sb0\rbrack+I\lbrack y,\,e\sb0\rbrack+
\frac{e_{0}}{2}m^{2},   \nonumber  \\ 
&&\Phi\lbrack y+x\sb{cl},\,e\sb0\rbrack
\rightarrow \Phi\lbrack
x\sb{cl},\,e\sb0\rbrack=\Psi(x\sb{out},\, x\sb{in},
\,e\sb{0}). 
\end{eqnarray}
Doing also a convenient replacement of variables 
$p\rightarrow\frac{p}{\sqrt e\sb0},\, y\rightarrow 
y\sqrt{e\sb0}$, we get
\begin{eqnarray}\label{e22}
&&S\sp{c}=\frac{i}{2}\int\sb0\sp\infty \frac{de\sb0}{e\sb0\sp2}\, 
\Psi(x\sb{out},\,x\sb{in},\,e\sb0)\, e\sp{{\rm i}I\lbrack
x\sb{cl},\,e\sb0
\rbrack}
\nonumber\\
&&\times\int\sb0\sp0 Dy \int Dp \exp\left\{\frac{i}{2}
\int\sb0\sp1 \left(p\sp{2}-\dot y\sp2-ge\sb0 y 
F\dot  y\right)d\tau\right\}\,.               
\end{eqnarray}
One can see that the path integral in (\ref{e22}) is, in fact, the
kernel of the Klein-Gordon propagator in the proper-time
representation.   
This path integral can be presented as
\begin{eqnarray*}
&&\int\sb0\sp0 D y \int Dp\exp\left\{\frac{i}{2}
\int\sb0\sp1 \left(p\sp{2}-\dot y\sp2-ge\sb0 y 
F\dot  y\right)d\tau\right\}= \\
&&\left\lbrack\frac{{\rm Det\,}
(\eta\sb{\mu\nu}\partial\sb\tau\sp2-{ge\sb0}F\sb{\mu\nu}\partial\sb\tau)}
{{\rm Det\,}(\eta\sb{\mu\nu}\partial\sb\tau\sp2)}\right\rbrack\sp{-1/2}
\int\sb0\sp0 D y \int Dp \exp\left\{\frac{i}{2}
\int\sb0\sp1 \left(p\sp{2}-\dot y\sp2\right)d\tau\right\}\,.
\end{eqnarray*}
Cancelling the factor ${\rm Det\,}(-\eta\sb{\mu\nu})$ in the ratio of
the determinants one obtains
\begin{equation}\label{e23}
\frac{{\rm Det\,}
(\eta\sb{\mu\nu}\partial\sb\tau\sp2-{ge\sb0}F\sb{\mu\nu}\partial\sb\tau)}
{{\rm Det\,}(\eta\sb{\mu\nu}\partial\sb\tau\sp2)} =
\frac{{\rm Det\,}
(-\delta\sp{\mu}\sb{\nu}\partial\sb\tau\sp2+{ge\sb0}F\sp{\mu}\sb{\quad\nu}
\partial\sb\tau)}{{\rm
Det\,}(-\delta\sp{\mu}\sb{\nu}\partial\sb\tau\sp2)}\,.  
\end{equation}
One can also make the replacement
\begin{equation}\label{e24}
-{\bf I}\partial\sb\tau\sp2+ge\sb0F\partial\sb\tau\rightarrow
-{\bf I}\partial\sb\tau\sp2+\frac{g\sp2e\sb0\sp2}{4}F\sp2\,,
\end{equation}
where {\bf I} stands for the unit 4$\times$4 matrix,
in the RHS of (\ref{e23}) because the spectra of both 
operators coincide. Indeed,
\begin{equation}\label{e25}
-{\bf I}\partial\sb\tau\sp2+ge\sb0F\partial\sb\tau=\exp\left(\frac 
{ge\sb0}{2}F\tau\right)\left(-{\bf I}\partial\sb\tau\sp2+
\frac{g\sp2e\sb0\sp2}{4}F\sp2\right)\exp\left(-\frac{ge\sb0}{2}F
\tau\right)\,,                                
\end{equation}
and  zero boundary conditions are invariant under the
transformation $ y\rightarrow \exp(\frac{ge\sb{0}F\tau}{2}) y$.
Then, using (\ref{e24}) and the value of the free path
integral \cite{b10},
\[
\frac{i}{2}\int\sb0\sp0 D y \int Dp \exp\{\frac{i}{2}\int 
d\tau(p\sp2-\dot y\sp2)\}
={1\over 8\pi\sp 2}\,,
\]
related, in fact, to the definition of the measure, we obtain
\begin{equation}\label{e26}
S\sp{c}=\frac{1}{8\pi\sp2}\int\sb0\sp\infty \frac{de\sb0}{e\sb0\sp2} 
\Psi(x\sb{out},\,x\sb{in},\,e\sb0)\, e\sp{{i}I\lbrack
x\sb{cl},\,e\sb0
\rbrack}
\left\lbrack\frac{{\rm Det\,}
(-{\bf I}\partial\sb\tau\sp2+\frac{g\sp2e\sb0\sp2}{4}F\sp2)}
{{\rm Det\,}(-{\bf I}\partial\sb\tau\sp2)}\right\rbrack\sp{-1/2}\,.
\end{equation}
The ratio of the determinants can be written now as 
\begin{eqnarray}\label{e27}
&&\frac{{\rm Det\,}
(-{\bf I}\partial\sb\tau\sp2+\frac{g\sp2e\sb0\sp2}{4}F\sp2)}
{{\rm Det\,}(-{\bf I}\partial\sb\tau\sp2)}=
\exp{\rm Tr\,}\left\lbrack \ln\left(-{\bf I}\partial\sb\tau\sp2+
\frac{g\sp2e\sb0\sp2}{4}F\sp2\right)-\ln \left(-{\bf I}\partial\sb
\tau\sp2\right)\right\rbrack                   \nonumber             \\
&&=\exp{\rm Tr}\left\lbrack\frac{e\sb0\sp2}{2}F\sp{2}\int\sb{0}\sp{g}
d\lambda\,\lambda\left(-{\bf I}\partial\sb\tau\sp2+
\frac{\lambda\sp2e\sb0\sp2}{4}F\sp2\right)\sp{-1}\right\rbrack
\nonumber                                                            \\
&&=\exp{\rm tr\,}\left\lbrack \frac{e\sb0\sp2}{2}F\sp{2}
\int\sb0\sp{g}d\lambda\,\lambda\sum\sb{n=1}\sp\infty
\left(\pi\sp2n\sp2{\bf I}+\frac{\lambda\sp2e\sb0\sp2}{4}F\sp2\right)
\sp{-1}\right\rbrack\,.                      
\end{eqnarray}                                    
The trace in the infinite-dimensional space in 
the second line of eq.(\ref{e27}) is taken
and only one  in the 4-dimensional space remains.
Using the formula
\[
\sum\sb{n=1}\sp\infty\left(\pi\sp2n\sp2+\kappa\sp2\right)\sp{-1}
=\frac{1}{2\kappa}\coth\kappa-\frac{1}{2\kappa\sp2}\,,
\]
which is also valid if $\kappa$ is an arbitrary $4\times 4$ matrix, and
integrating in (\ref{e27}), we find
\begin{equation}
\label{e28}
\frac{{\rm Det\,}
(-{\bf I}\partial\sb\tau\sp2+\frac{g\sp2e\sb0\sp2}{4}F\sp2)}
{{\rm Det\,}(-{\bf I}\partial\sb\tau\sp2)}= \det\left(\frac{\sinh\frac
{ge\sb{0}F}{2}}{\frac{ge\sb{0}F}{2}}\right)\,.
\end{equation} 
Thus,
\begin{equation}
\label{e29}
S\sp c={1\over 32\pi\sp 2}\int\sb 0\sp\infty de\sb 0 
\left(\det\frac{\sinh\frac{e\sb0gF}{2}}{gF}\right)\sp
{-\frac{1}{2}}\Psi(x\sb{out},\,x\sb{in},\,e\sb0)e\sp{{i}I\lbrack x\sb{cl},
e\sb0\rbrack}\,,                                             
\end{equation}
where the function $\Psi(x\sb{out},\,x\sb{in},\,e\sb0)$ is SF 
on the classical trajectory $x\sb{cl}$. The latter can be
easily found solving the eq.(\ref{e20}):
\begin{equation}
\label{e30}
x\sb{cl}=\left(\exp(ge\sb0F)-\eta\right)\sp{-1}\left\lbrack
\exp(ge\sb0F\tau)(x\sb{out}
-x\sb{in})+\exp(ge\sb0F)x\sb{in}-x\sb{out}\right\rbrack\,.
\end{equation}
Substituting (\ref{e30}) into  eqs.(\ref{e21}) and (\ref{e29}), we obtain
\begin{eqnarray}
\label{e31}
&&S\sp c={1\over 32\pi\sp 2}\int\sb 0\sp\infty de\sb 0 
\left(\det\frac{\sinh\frac{ge\sb0F}{2}}{gF}\right)\sp
{-\frac{1}{2}}\Psi(x\sb{out},\,x\sb{in},\,e\sb0)            \nonumber \\
&&\times\exp\left\{\frac{{i}g}{2}x\sb{out}Fx\sb{in}-
\frac{i}{2}e\sb0m\sp 2-\frac{{i}g}{4}(x\sb{out}-
x\sb{in})F\coth\left(\frac{ge\sb 0F}{2}\right)(x\sb{out}-x\sb{in})\right\}.
\end{eqnarray}
where 
\begin{eqnarray}\label{e32}
&&\Psi(x\sb{out},x\sb{in},e\sb 0)
=\left\lbrack m+\frac{g}{2}(x\sb{out}-x\sb{in})F
\left(\coth\frac{ge\sb0F}{2}-1\right)\gamma\right\rbrack    \nonumber  \\
&\times&\sqrt{\det\cosh\frac{ge\sb{0}F}{2}}
\left [1-\frac{i}{2}
\left(\tanh\frac{ge\sb{0}F}{2}\right)\sb{\mu\nu}\sigma\sp{\mu\nu}\right.
\nonumber \\
&+&\left.\frac{1}{8}\epsilon\sp{\alpha\beta\mu\nu}
\left(\tanh\frac{ge\sb{0}F}{2}\right)\sb{\alpha\beta} 
\left(\tanh\frac{ge\sb{0}F}{2}\right)\sb{\mu\nu}\gamma
\sp5\right\rbrack\,.                       
\end{eqnarray}
Now we are going to compare the representation (\ref{e31}) with
the  Schwinger formula \cite{b14}, which he has been derived in the same
case of constant field by means of the proper-time method. 
The Schwinger representation has the form 
\begin{eqnarray}
\label{e33}
&&S\sp{c}(x\sb{out},\,x\sb{in})=\frac{1}{32\pi\sp2}\left\lbrack
\gamma\sp{\mu}
\left({i}\frac{\partial}{\partial x\sb{out}\sp{\mu}}-
gA\sb{\mu}(x\sb{out})\right)+m\right\rbrack\int\sb{0}\sp{\infty}de\sb{0}
\left(\det\frac{\sinh\frac{ge\sb{0}F}{2}}{gF}\right)\sp{-1/2}
                                                  \nonumber          \\
&&\times
\exp\left\{\frac{i}{2}\left\lbrack gx\sb{out}Fx\sb{in}-e\sb{0}
m\sp{2}-(x\sb{out}-x\sb{in})\frac{gF}{2}\coth\frac{ge\sb{0}F}{2}
(x\sb{out}-x\sb{in})-\frac{ge\sb{0}}{2}F\sb{\mu\nu}\sigma\sp{\mu\nu}
\right\rbrack\right\}.                            
\end{eqnarray}
Doing the differentiation with respect to $x\sb{out}\sp\mu$ we
transform the formula (\ref{e33}) to a form which is convenient for
the comparison with our representation (\ref{e29}),
\begin{eqnarray}\label{e34}
&&S\sp c=\frac{1}{32\pi\sp 2}\int\sb 0\sp\infty 
de\sb0\left(\det\frac{\sinh
\frac{ge\sb0F}{2}}{gF}\right)\sp{-1/2} 
\Psi\sb{S}(x\sb{out}, x\sb{in}, e\sb 0) \nonumber \\
&&\times\exp\left\{{i}\frac{g}{2}x\sb{out}Fx\sb{in}-\frac{i}{2}
e\sb 0m\sp 2-{i}\frac{g}{4}(x\sb{out}-x\sb{in})F\coth
\left(\frac{ge\sb0F}{2}\right)(x\sb{out}-x\sb{in})\right\}\,,
\end{eqnarray}
where the function $\Psi\sb{S}$ is given by 
\begin{equation}\label{e35}
\Psi\sb{S}(x\sb{out},x\sb{in},e\sb 0)=
\left\lbrack m+\frac{g}{2}(x\sb{out}-x\sb{in})F
(\coth\frac{ge\sb0F}{2}-1)\gamma\right\rbrack
\exp(-{i}\frac{e\sb 0g}{4}F\sb{\mu\nu}\sigma\sp{\mu\nu})\,.     
\end{equation}
Thus one needs only to compare the functions $\Psi$ and $\Psi\sb{S}$.
They coincide, since the following formula takes place (see Appendix
B), where
$\omega\sb{\mu\nu}$ is an arbitrary antisymmetric tensor,
\begin{eqnarray}
\label{e36}
\exp\left(-\frac{i}{4}\omega_{\mu\nu}\sigma ^{\mu\nu}\right)&=&
\sqrt{\det\cosh\frac{\omega}{2}}\left [
1-\frac{i}{2}
\left(\tanh\frac{\omega}{2}
\right)\sb{\mu\nu}\sigma\sp{\mu\nu}\right. \nonumber \\
&+&\left.\frac{1}{8}\epsilon\sp{\alpha\beta\mu\nu}
\left(\tanh\frac{\omega}{2}\right)\sb{\alpha\beta} 
\left(\tanh\frac{\omega}{2}\right)\sb{\mu\nu}\gamma\sp5\right\rbrack\,.
\end{eqnarray}
In fact, the latter formula presents a linear decomposition of a finite
Lorentz transformation in the independent $\gamma$-matrix structures.

\subsection{Propagator in a  constant uniform field and a plane wave field}

The 4-potential
\begin{equation}
\label{e101}
A\sb{\mu}\sp{comb}=-\frac{1}{2}F\sb{\mu\nu}x\sp{\nu}+a\sb{\mu}(nx),
\end{equation}
where $a\sb{\mu}(\phi)$ is a vector-valued function of a 
real variable $\phi$ and $n$ is a normalized isotropic vector
$n^{\mu}=(1,\,{\bf n})$,
\begin{equation}
\label{e102}
n\sp{2}=0,\qquad {\bf n}\sp{2}=1,
\end{equation}
produces the field
\begin{equation}
\label{e103}
F\sb{\mu\nu}\sp{comb}(nx)=F\sb{\mu\nu}+f\sb{\mu\nu}(nx),
\end{equation}
which is a superposition of the constant field $F\sb{\mu\nu}$ and the 
plane-wave field
\[
f\sb{\mu\nu}(nx)=n\sb{\mu}a\sp{\prime}\sb{\nu}(nx)-n\sb{\nu}
a\sp{\prime}\sb{\mu}(nx)
\]
Without loss of generality we may choose $a\sb{\mu}$ to be transversal,
\begin{equation}
\label{e105a}
n\sp{\mu}a\sb{\mu}(\phi)=0.
\end{equation}  

The dependence of SF $\Phi\lbrack x,e\sb{0}\rbrack$ on the
trajectory $x\sp{\mu}(\tau)$ is twofold. In addition to the direct dependence
(see eq.(\ref{e13})), there is an indirect one through the external 
field. In the
case under consideration the field depends on $x\sb{\mu}(\tau)$ 
only through the scalar combination $nx(\tau)$. Replacing
the latter by an auxiliary scalar trajectory $\phi(\tau)$ one obtains
\begin{eqnarray}
\label{e106}
&&\tilde\Phi\lbrack x,\phi, e\sb{0}\rbrack=\left\lbrack m+(2e\sb{0})\sp{-1}
\dot{x}\sp{\mu}\star\tilde{K}\sb{\mu\lambda}\left(2\eta\sp{\lambda\kappa}
-ge\sb{0}\tilde{B}\sp{\lambda\kappa}\right)\gamma\sb{\kappa}
\right. \nonumber                                                \\
&&-\left.\frac{ig}{4}\left(me\sb{0}+\dot{x}\sp{\mu}\star
\tilde{K}\sb{\mu\lambda} 
\gamma\sp{\lambda}\right)\tilde{B}\sb{\kappa\nu}\sigma\sp{\kappa\nu}+
m\frac{g\sp{2}e\sb{0}\sp{2}}{16}\tilde{B}\sb{\alpha\beta}
\tilde{B}\sp{\ast\alpha\beta}
\gamma\sp{5}\right\rbrack  \nonumber                              \\
&&\times\exp\left\{-\frac{e\sb{0}}{2}\int\sb{0}\sp{g}dg\sp{\prime}
{\rm Tr}\tilde
{{\cal R}}(g\sp{\prime})\star{\cal F}\sp{comb}(\phi)\right\},
\end{eqnarray}
where  ${\cal F}\sb{\mu\nu}\sp{comb}(\phi\vert\tau-\tau\sp{\prime})=
\lbrack F\sb{\mu\nu}+f\sb{\mu\nu}(\phi(\tau))\rbrack\delta(\tau,
\tau\sp{\prime})$ and
\begin{eqnarray}
\label{e107}
&&\tilde{B}\sb{\mu\nu}=F\sb{\mu\lambda}\sp{comb}(\phi)\star\tilde{K}
\sp{\lambda}\sb{\quad\nu},\quad
\tilde{K}\sb{\mu\nu}=\eta\sb{\mu\nu}+ge\sb{0}\tilde{{\cal R}}\sb{\mu\nu}(g)
\star F\sp{comb}(\phi),\\
&&\label{e107b}
\left[\eta\frac{\partial}{\partial\tau}-ge\sb{0}F\sp{comb}(\phi(\tau))
\right]\tilde{{\cal R}}(q\vert\tau,\tau\sp{\prime})=\eta\delta(\tau,\tau
\sp{\prime}), \\
&&\label{e107c}
\tilde{{\cal R}}(g\vert1,\tau)=-\tilde{\cal R}
(g\vert0,\tau),\quad \forall \tau\in(0,1).
\end{eqnarray}  
Obviously,
\begin{equation}
\label{e107a}
\left.\tilde{{\cal R}}(g)\right\vert\sb{\phi(\tau)=nx(\tau)}=
{\cal R}(g),\quad
\left.\tilde{K}\right\vert\sb{\phi(\tau)=nx(\tau)}=K,\quad
\left.\tilde{B}\right\vert\sb{\phi(\tau)=nx(\tau)}=B,
\end{equation}
and, therefore, 
\begin{equation}
\label{e108}
\tilde{\Phi}\lbrack x,nx,e\sb{0}\rbrack=\Phi\lbrack x,e\sb{0}\rbrack.
\end{equation}
Inserting the integral of a $\delta$-function,
\[
\int D\phi\,D\lambda\;e\sp{i\lambda\star(\phi-nx)}=1,
\]
into the RHS of eq.(\ref{e10}) and using (\ref{e108}) one transforms 
the path integral (\ref{e10}) into a quasi-Gaussian one of simple form
\begin{eqnarray}
\label{e108a}
&&S\sp{c}(x\sb{out},x\sb{in})=\frac{i}{2}\int\sb{0}\sp{\infty}
de\sb{0}\int D\phi\, D\lambda\, e\sp{i\lambda\star\phi}
\int\sb{x\sb{in}}\sp{x\sb{out}}Dx\,M(e\sb{0})\tilde{\Phi}
\lbrack x,\phi,e\sb{0}\rbrack\nonumber   \\
&&\times\exp\{i\tilde{I}\lbrack x,\phi,e\sb{0}
\rbrack-i\lambda\star(nx)\}.
\end{eqnarray}  
The action functional 
\begin{equation}
\label{e109}
\tilde{I}\lbrack x,\phi,e\sb{0}\rbrack=-\frac{1}{2e\sb{0}}\dot{x}\star
\dot{x}-\frac{e\sb{0}}{2}m\sp{2}-\frac{g}{2}x\star\bar{\cal F}\star\dot{x}
-ga(\phi)\star\dot{x}
\end{equation}
(where $\bar{\cal F}(\tau,\tau^{\prime})=F\delta(\tau-\tau^{\prime})$)
contains only linear and bilinear terms in $x$ (and the 
bilinear part does not depend on the wave potential $a\sb{\mu}$).
SF $\tilde\Phi\lbrack x,\phi,e\sb{0}\rbrack$ is linear in $x$
and (following the same way of reasoning as in the case of a
constant field) one finds
\begin{equation}
\label{e110}
S\sp{c}=\frac{1}{32\pi\sp{2}}\int\sb{0}\sp{\infty}de\sb{0}
\left(\det\frac{\sinh\frac{ge\sb{0}F}{2}}{gF}\right)\sp{-\frac{1}{2}}
\int D\phi\, D\lambda\; e\sp{i\lambda\star(\phi-nx\sb{q})}
\nonumber                                               \\
\tilde{\Phi}\lbrack x\sb{q}, \phi,e\sb{0}\rbrack
e^{i\tilde{I}\lbrack x\sb{q},\phi,e\sb{0}\rbrack},
\end{equation}
where $x\sb{q}$ is the solution to the equation
\begin{equation}
\label{e111}
\ddot{x}_{q}-ge_{0}F\dot{x}_{q}=e_{0}\lambda{n}-e_{0}ga^{\prime}
(\phi)\dot\phi\,,
\end{equation}
obeying the boundary conditions 
\begin{equation}
\label{e112}
x\sb{q}(0)=x\sb{in},\qquad x\sb{q}(1)=x\sb{out}.        
\end{equation}
Introducing an appropriate
Green function ${\cal G}={\cal G}(\tau,\tau\sp{\prime})$
for the second-order operator,
\begin{eqnarray}
\label{e113}
&&\left\lbrack\eta\frac{\partial\sp{2}}{\partial\tau\sp{2}}
-ge\sb{0}F\frac{\partial}{\partial\tau}\right\rbrack 
{\cal G}(\tau,\tau\sp{\prime})=\eta\delta(\tau-\tau\sp{\prime}),\\
&&{\cal G}(0,\tau)={\cal G}(1,\tau)={\cal G}(\tau,0)={\cal G}(\tau,1)=0, 
\quad\forall\tau\in(0,1),
\end{eqnarray}
one  presents this solution in the form
\begin{equation}
\label{e114}
x\sb{q}=x\sb{cl}+e\sb{0}{\cal G}\star\left(\lambda n-ga\sp{\prime}
(\phi)\dot{\phi}\right).
\end{equation}
The value of the action functional $\tilde{I}[x,\phi,e_{0}]$ on the 
solution $x_{q}$ is given by
\begin{eqnarray}
\label{e115}
&&\tilde{I}[x_{q},\phi,e_{0}]=\bar{I}[x_{cl},e_{0}]-ga(\phi) 
\star\dot x_{cl}                          \nonumber    \\
&&-\frac{e_{0}}{2}\left(ga^{\prime}(\phi)\dot\phi-\lambda n\right)
\star{\cal G}\star\left(ga^{\prime}(\phi)\dot\phi-\lambda n\right)
+\lambda{n}\star(x_{q}-x_{cl}),
\end{eqnarray}
where
\begin{equation}
\bar{I}[x,e_{0}]=-\frac{1}{2e_{0}}\dot{x}\star\dot{x}-\frac{e_{0}}
{2}m^{2}-\frac{g}{2}x\star\bar{\cal F}\star\dot{x}
\end{equation}
is the action in a uniform constant field $F$.

The functional integral over $\lambda$ in  (\ref{e108a}) is a 
quasi-Gaussian one of simple form
(let us remind that $x_{q}$ is linear in $\lambda$, 
see eq.(\ref{e114})) and the integration can be done explicitly. 
The result is a 
formula for the propagator in which the only functional integration is 
over the scalar trajectory $\phi(\tau)$. However, the latter integration
is hardly to be performed explicitly in the general case (for
arbitrary $a_{\mu}(\phi)$). Nevertheless, there exists a specific
combination \cite{b3} for which the 
integration can be done and explicit formulae for the propagator to be 
derived. The latter are comparable with the corresponding Schwinger-type
formulae \cite{b9}, which are also explicit in this case.

Namely, let us choose the wave vector $n$ to coincide with a real eigenvector
of the matrix $F$ (see Appendix A),
\begin{equation}
\label{e115a}
F_{\mu\nu}n^{\nu}=-{\cal E}n_{\mu}, \qquad n^{2}=0, \qquad 
{\bf n}^{2}=1,
\end{equation}
In this case $nx_{q}=nx_{cl}$, and, moreover,
the action functional (\ref{e109}) is ``on-shell'' invariant with
respect to longitudinal shifts
\begin{equation}
\label{e116}
x_{q}(\tau)\rightarrow x_{q}(\tau)+\alpha(\tau)n,
\qquad 
\alpha(0)=\alpha(1)=0,
\end{equation}
by virtue of (\ref{e115a}) and the transversality (\ref{e105a}) of the 
wave potential $a_{\mu}$.
Then $\tilde{I}[x_{q},\phi,e_{0}]$ does not depend on $\lambda$,
\begin{equation}
\label{e116a}
\tilde{I}[x_{q},\phi,e_{0}]=\tilde{I}[x_{tr},\phi,e_{0}]
\end{equation}
where
\begin{equation}
\label{e116b}
x_{tr}=x_{cl}-ge_{0}{\cal G}\star\left(a^{\prime}(\phi)\dot\phi\right)
\end{equation}
is a  solution to the equation
\begin{equation}
\label{e117}
\ddot{x}_{tr}-ge_{0}F\dot{x}_{tr}=-ge_{0}a^{\prime}(\phi)\dot{\phi}\,,
\end{equation}
obeying the boundary conditions (\ref{e111}).
However SF $\tilde\Phi[x_{q},\phi,e_{0}]$ does not show
this invariance
and, therefore, is $\lambda$-dependent. 
Presenting $x_{q}$ as a sum $x_{q}=x_{tr}+e_{0}{\cal G}n\star\lambda$,
and substituting it into  the expansion of SF in the 
antisymmetrized  products of $\gamma$-matrices,
\begin{eqnarray}
\label{e120}
&&\tilde\Phi[x,\phi,e_{0}]=\left[e_{0}^{-1}\dot{x}^{\mu}\star\tilde{K}
_{\mu\nu}\left(\gamma^{\nu}+\frac{ge_{0}}{4}\tilde{B}_{\alpha\beta}
\gamma^{[\nu}\gamma^{\alpha}\gamma^{\beta]}\right)\right. \nonumber\\ 
&&\left.+m\left(1+\frac{ge_{0}}{4}\tilde{B}_{\alpha\beta}\gamma^{[\alpha}
\gamma^{\beta]}+\frac{g^{2}e^{2}_{0}}{32}\tilde{B}_{\alpha\beta}
\tilde{B}_{\mu\nu}\gamma^{[\alpha}\gamma^{\beta}\gamma^{\mu}\gamma^{\nu]}
\right)\right]\Lambda[\phi,e_{0}],          \\
\label{e120a}
&&\Lambda[\phi,e_{0}]=\exp\left\{-\frac{e_{0}}{2}\int_{0}^{g}
dg^{\prime}\Tr\tilde{\cal R}(g^{\prime})\star{\cal F}^{comb}(\phi)\right\}.
\end{eqnarray}
we obtain,
\begin{eqnarray}
\label{e121}
&&\tilde{\Phi}[x_{q},\phi,e_{0}]=\tilde{\Phi}[x_{tr},\phi,e_{0}]+
\lambda\star{l}[\phi,e_{0}],   \\
\label{e122}
&&l[\phi,e_{0}]=n^{\kappa}{\cal G}^{(r)}_{\kappa\mu}\star 
\tilde{K}^{\mu}_{\quad\nu}\left(\gamma^{\nu}+\frac{ge_{0}}{4}
\tilde{B}_{\alpha\beta}\gamma^{[\nu}\gamma^{\alpha}\gamma^{\beta]}
\right){\Lambda}[\phi,e_{0}],   \\
&&{\cal G}^{(r)}(\tau,\tau^{\prime})=\frac{\partial}
{\partial\tau^{\prime}}
{\cal G}(\tau,\tau^{\prime}).
\end{eqnarray}
It turns out that ${l}[\phi,e_{0}]$ does not depend on $\phi$.
First, expanding $\tilde{\cal R}$ in powers of $f$ and using (\ref{e115a})
and (\ref{e105a}) one derives that ${\Lambda}[\phi,e_{0}]$ coincides
with the expression
\begin{equation}
\label{e123}
\bar{\Lambda}(e_{0})=\exp\left\{\frac{e_{0}}{2}\int_{O}^{g}dg^{\prime}
\Tr\bar{\cal R}(g)\star\bar{\cal F}\right\}.
\end{equation} 
Second\footnote{In this section we denote by $\bar{\cal R}(g),\,
\bar{K},\, \bar{B}$ the quantities given by (\ref{e17}), i.e.,
corresponding to the case of a constant uniform field $F$.},
\begin{equation}
\label{e124}
n_{\mu}\tilde{K}^{\mu}_{\quad \nu}= n_{\mu}\bar{K}^{\mu}_{\quad \nu}=
\frac{1}{2}\left(n\bar{K}\bar{n}\right)n_{\nu}.
\end{equation}
Indeed, using the definitions  (\ref{e107a}) one finds that 
$\tilde{K}$ satisfies the equation
\begin{equation}
\label{e125}
\left[\frac{\partial}{\partial \tau}-ge_{0}(F+f(\phi(t)))\right]
\tilde{K}(\tau)=0,
\end{equation}
and the boundary conditions
\begin{equation}
\label{e126}
\tilde{K}(0)+\tilde{K}(1)=2\eta.
\end{equation}
Multiplying (\ref{e125}) by $n$ and using the properties
(\ref{e115a}), (\ref{e105a}) we find
\begin{equation}
\label{e126a}
\left(\frac{\partial}{\partial\tau}-ge_{0}{\cal E}\right)n\tilde{K}=0,
\qquad n\tilde{K}(0)+n\tilde{K}(1)=2n.
\end{equation}
At the same time $n\bar{K}$ obeys (\ref{e126a}).
Therefore, $n\tilde{K}$ and $n\bar{K}$ coincide. Then using
(\ref{e105a}) and the properties of $n, \bar{n}$ (see Appendix A)
one gets (\ref{e124}).
Third, using the same properties of the electromagnetic field one 
can derive
\begin{equation}
\label{e127}
\tilde{B}_{\alpha\beta}=\bar{B}_{\alpha\beta}+n_{\alpha}b_{\beta}-
b_{\alpha}n_{\beta},
\end{equation}
where $b_{\alpha}$ depends on $\phi$ and $\bar{B}$ is given by (\ref{e17}).
Substituting (\ref{e127}) into (\ref{e122}) and using (\ref{e115a})
one finds
\begin{eqnarray*}
&&{l}[\phi,e_{0}]=\nonumber \\
&&\frac{1}{4}\left(n{\cal G}^{(r)}\bar{n}\right)
\star\left(n\bar{K}\bar{n}\right)\left[n_{\nu}\gamma^{\nu}+
\frac{ge_{0}}{4}n_{\nu}\left(\bar{B}_{\alpha\beta}+n_{\alpha}b_{\beta}
-b_{\alpha}n_{\beta}\right)\gamma^{[\nu}\gamma^{\alpha}\gamma^{\beta]}
\right]\bar{\Lambda}(e_{0}).
\end{eqnarray*}
and the contribution of the $\phi$-dependent terms vanishes by virtue
of the complete antisymmetry of $\gamma^{[\nu}\gamma^{\alpha}
\gamma^{\beta]}$. Therefore ${l}[\phi,e_{0}]$ can be replaced
by
\begin{equation}
\label{e127a}
\bar{l}(e_{0})=n^{\kappa}{\cal G}^{(r)}_{\kappa\mu}\star 
\bar{K}^{\mu}_{\quad\nu}\left(\gamma^{\nu}+\frac{ge_{0}}{4}
\bar{B}_{\alpha\beta}\gamma^{[\nu}\gamma^{\alpha}\gamma^{\beta]}
\right)\bar{\Lambda}(e_{0}).
\end{equation}
Substituting (\ref{e127a}) into (\ref{e121}) and then into
(\ref{e110}), using (\ref{e116a}) and
\[
\lambda(\tau)e^{i\lambda\star\phi}=-i\frac{\delta}{\delta\phi
(\tau)}e^{i\lambda\star\phi},
\]
and integrating by parts we find
\begin{eqnarray}
\label{e128}
&&S\sp{c}=\frac{1}{32\pi\sp{2}}\int\sb{0}\sp{\infty}de\sb{0}
\left(\det\frac{\sinh\frac{ge\sb{0}F}{2}}{gF}\right)\sp{-\frac{1}{2}}
\int D\phi\, D\lambda\; e\sp{i\lambda\star(\phi-nx\sb{cl})}
 \nonumber                                               \\
&&\times\left[\tilde{\Phi}[ x\sb{tr}, \phi,e\sb{0}]-\left(\frac{\delta}
{\delta\phi}\tilde{I}[x_{tr},\phi,e_{0}]\right)\star\bar{l}(e_{0})\right]
\exp\left\{i\tilde{I}\lbrack x\sb{tr},\phi,e\sb{0}\rbrack\right\}.
\end{eqnarray}
Inserting the derivative,
\[
\frac{\delta}{\delta\phi}\tilde{I}[x_{tr},\phi,e_{0}]=
-a^{\prime}(\phi)\dot{x}_{tr},
\]
and using (\ref{e127}), (\ref{e122}) we transform (\ref{e128}) to 
the following form
\begin{eqnarray}
\label{e129}
&&S\sp{c}=\frac{1}{32\pi\sp{2}}\int\sb{0}\sp{\infty}de\sb{0}
\left(\det\frac{\sinh\frac{ge\sb{0}F}{2}}{gF}\right)\sp{-\frac{1}{2}}
\int D\phi\, D\lambda\; e\sp{i\lambda\star(\phi-nx\sb{cl})}
  \nonumber                                               \\
&&\times\tilde{\Phi}\lbrack \tilde{x}, \phi,e\sb{0}\rbrack
\exp\left\{i\tilde{I}\lbrack x\sb{tr},\phi,e\sb{0}\rbrack\right\},
\end{eqnarray}
where
\[
\tilde{x}^{\mu}=x^{\mu}_{tr}+ge_{0}n^{\mu}a^{\prime}_{\kappa}
(\phi)\dot{x}^{\kappa}_{tr}.
\]
One can straightforwardly check that $\tilde{x}$ satisfies the equation
\begin{equation}
\label{e130}
\ddot{\tilde{x}}_{\mu}-ge_{0}\left(F_{\mu\nu}+n_{\mu}a^{\prime}_{\nu}
(\phi)\right)\dot{\tilde{x}}^{\nu}=-ge_{0}a^{\prime}_{\mu}(\phi)
\dot{\phi}
\end{equation}
and the boundary conditions (\ref{e111}). The trajectory $x_{tr}$ in the
action $\tilde{I}$ can be replaced by $\tilde{x}$ due to the invariance 
of the action $\tilde{I}[x,\phi,e_{0}]$ under the longitudinal shifts 
(\ref{e116}). The integration over $\lambda$ and $\phi$ is straightforward
now. One needs only to take into account that
$\tilde{x}|_{\phi=nx_{cl}}\equiv{x}_{comb}$
is the solution (subjected to the boundary conditions (\ref{e111})) 
to the equation of  motion 
\begin{equation}
\label{e131}
\ddot{x}^{\mu}_{comb}-ge_{0}\left(F+f(nx_{comb})\right)^{\mu}_{\quad\nu}
\dot{x}^{\nu}_{comb}=0.
\end{equation}
Indeed, eq.(\ref{e130}) turns out to be equivalent to eq.(\ref{e131}) when
$\phi=nx_{cl}$ and the relation $nx_{comb}=nx_{cl}$
is taken into account. Therefore,
\begin{equation}
\label{e132}
\left.\tilde{\Phi}[\tilde{x},\phi,e_{0}]\right|_{\phi=nx_{cl}}=
\Phi[x_{comb},e_{0}],
\qquad
\left.\tilde{I}[\tilde{x},\phi,e_{0}]\right|_{\phi=nx_{cl}}
=I[x_{comb},e_{0}].
\end{equation}
Finally, we get
\begin{equation}
\label{e133}
S^{c}=\frac{1}{32\pi^{2}}\int_{0}^{\infty}de_{0}\left(\det\frac{\sinh\frac
{ge_{0}F}{2}}{gF}\right)^{-\frac{1}{2}}\Phi[x_{comb},e_{0}]e^{I[x_{comb},e_{0}]},
\end{equation}
where
\begin{eqnarray}
\label{e133a}
&&\Phi[x_{comb},e_{0}]=\left[e_{0}^{-1}\dot{x}^{\mu}_{comb}\star{K}_{\mu\nu}
\left(\gamma^{\nu}+\frac{ge_{0}}{4}B_{\alpha\beta}\gamma^{[\nu}
\gamma^{\alpha}
\gamma^{\beta]}\right)\right.\nonumber  \\
&&\left.+m\left(1+\frac{ge_{0}}{4}B_{\alpha\beta}\gamma^{[\alpha}\gamma^
{\beta]}+\frac{g^{2}e^{2}_{0}}{32}B_{\alpha\beta}B_{\mu\nu}
\gamma^{[\alpha}\gamma^{\beta}\gamma^{\mu}\gamma^{\nu]}\right)\right]
\bar{\Lambda}(e_{0}).
\end{eqnarray}
The vector $\dot{x}_{comb}$ satisfies (\ref{e131}) and can be
presented as
\begin{equation}
\label{e134}
\dot{x}_{comb}(\tau)=T_{A}\exp\left\{-ge_{0}\int_{\tau}
^{1}F^{comb}(nx_{cl}(\tau))d\tau\right\}\dot{x}(1),
\end{equation}
where $T_{A}$ denotes the antichronological product.
On the other hand the tensor trajectory $K(\tau)$ satisfies
(\ref{e125}) where one has to replace $\phi$ by $nx_{cl}$.
Therefore
\begin{equation}
K(\tau)=T_{A}\exp\left\{-ge_{0}\int_{\tau}^{1}F^{comb}(nx_{cl})d\tau
\right\}K(1),
\end{equation}
and\footnote{The operator
$T_{A}\exp\left\{-ge_{0}\int_{\tau}^{1}F^{comb}
(nx_{cl})d\tau\right\}$ preserves the scalar product due to the
antisymmetry of the stress tensor.}
\begin{equation}
\label{e134a}
\dot{x}_{comb}\star{K}=\dot{x}_{comb}(1)K(1).
\end{equation}
Substituting (\ref{e134a}) into (\ref{e133a}) and taking into account the 
relation 
$B_{[\alpha\beta}B_{\mu\nu]}=\bar{B}_{[\alpha\beta}\bar{B}_{\mu\nu]}$
we find
\begin{eqnarray}
\label{e135}
&&\Phi[x_{comb},e_{0}]=\left[e_{0}^{-1}\dot{x}_{comb}^{\mu}
(1)K_{\mu\nu}(1)
\left(\gamma^{\nu}+\frac{ge_{0}}{4}B_{\alpha\beta}\gamma^{[\nu}
\gamma^{\alpha}\gamma^{\beta]}\right)\right.\nonumber \\
&&+m\left.\left(1+\frac{ge_{0}}{4}B_{\alpha\beta}\gamma^{[\alpha}
\gamma^{\beta]}+\frac{g^{2}e_{0}^{2}}{32}\bar{B}_{\alpha\beta}
\bar{B}_{\mu\nu}\gamma^{[\alpha}\gamma^{\beta}\gamma^{\mu}\gamma^{\nu]}
\right)\right]\bar{\Lambda}(e_{0}).
\end{eqnarray}

A representation for the propagator in this specific field combination 
(characterized by the relation (\ref{e115a})) was given \cite{b9a,b9} in terms of
proper-time integral only. Another more complicated representation has been
obtained before in \cite{b3}.
In our notation the representation \cite{b9a,b9}  can be 
written as 
\begin{eqnarray}
\label{e136}
&&S^{c}(x_{out},x_{in})=\left[\gamma^{\mu}
\left(i\frac{\partial}{\partial{x}^{\mu}_{out}}-gA^{comb}_
{\mu}(nx_{out})\right)+m\right]       \nonumber   \\
&&\times\frac{1}{32\pi^{2}}\int_{0}^{\infty}de_{0}
\left(\det\frac{\sinh\frac{ge_{0}F}{2}}{gF}\right)^{-\frac{1}{2}}
e^{iI[x_{comb},e_{0}]}\Delta[nx_{cl},e_{0}],
\end{eqnarray}
where
\begin{eqnarray}
\label{e137}
&&\Delta[nx_{cl},e_{0}]=T\exp\left\{-\frac{ige_{0}}{4}\int_{0}^{1}
d\tau\left(F+f(nx_{cl}(\tau))\right)_{\mu\nu}\sigma^{\mu\nu}\right\}\\
\label{e138}
&&=\exp\left\{-\frac{ige_{0}}{4}F_{\mu\nu}
\sigma^{\mu\nu}\right\}-\frac{ige_{0}}{4}\int_{0}^{1}d\tau
\left(e^{\frac{ge_{0}}{2}F(1-2\tau)}
f(nx_{cl}(\tau))e^{-\frac{ge_{0}}{2}F(1-2\tau)}\right)_{\mu\nu}
\sigma^{\mu\nu},
\end{eqnarray}
and  $T$ denotes chronological product.
Taking the derivative in (\ref{e136}) one can use the relation
\begin{equation}
\label{e139}
\frac{\partial}{\partial{x}^{\mu}_{out}}I[x_{comb},e_{0}]=-p_{\mu}(1),
\end{equation}
where 
\[
p_{\mu}(\tau)=-
\left.\frac{\delta}{\delta\dot{x}^{\mu}}I[x,e_{0}]
\right\vert_{x=x_{comb}},
\]
is the on-shell momentum, in particular,
\begin{equation}
p(1)=e^{-1}_{0}\dot{x}_{comb}(1)+ge_{0}A^{comb}(nx_{out}).
\end{equation}
On the other hand,
\begin{equation}
\label{e140}
\gamma^{\mu}\frac{\partial}{\partial{x}^{\mu}_{out}}\Delta
[nx_{cl},e_{0}]=0.
\end{equation}
Indeed, one gets from eq.(\ref{e30}, with the aid of
eq.(\ref{e115a}),
\begin{equation}
\label{e141}
\gamma^{\mu}\frac{\partial}{\partial{x}^{\mu}_{out}}\left(nx_{cl}
(\tau)\right)=\frac{e^{ge_{0}{\cal E}\tau}-1}{e^{ge_{0}{\cal E}}-1}
n_{\mu}\gamma^{\mu}.
\end{equation}
Then, using the representation (\ref{e138}) for
$\Delta[nx_{cl},e_{0}]$, eqs.(\ref{e102}) and (\ref{e105a}), and the 
properties of the $\gamma$-matrices, one easily derives (\ref{e140}).
Differentiating in (\ref{e136}), one obtains,  with the aid of 
(\ref{e138}), (\ref{e139}) and (\ref{e140}),
\begin{equation}
\label{e142}
S^{c}(x_{out},x_{in})=\frac{1}{32\pi^{2}}\int_{0}^{\infty}de_{0}
\left(\det\frac{\sinh\frac{ge_{0}F}{2}}{gF}\right)^{-\frac{1}{2}}
\Psi^{comb}_{S}(x_{out},x_{in}, e_{0})e^{iI[x_{comb},e_{0}]},
\end{equation}
where
\[
\Psi_{S}^{comb}(x_{out},x_{in},e_{0})=
\left(e_{0}^{-1}\dot{x}^{\mu}_{comb}(1)\gamma_{\mu}+m\right)
\Delta[nx_{cl},e_{0}].
\]
Using the identities (\ref{a3-02}) and (\ref{a3-03})
one can verify that
\[
\Psi_{S}^{comb}(x_{out},x_{in},e_{0})=\Phi[x_{comb},e_{0}].
\]
Thus the representations (\ref{e133}) and (\ref{e136}) are equivalent.

\section{Spin factor in $2+1$ dimensions}

\subsection{Derivation of spin factor}

In $2+1$ dimensions the equation for the Dirac  propagator has the
form 
\begin{equation}
\label{d1}
\left[\gamma^{\mu}\left(i\partial_{\mu}-gA_{\mu}(x)\right)-m\right]S^{c}(x,y)
=-\delta^{3}(x-y)\;,\;
\end{equation}
where $\gamma$ matrices in $2+1$ dimensions can be
taken, for example, in the form
$\gamma^0=\sigma^3,\;\gamma^1=i\sigma^2,\;\gamma^2=-i\sigma^1,\;
\left[\gamma^{\mu}\;,\;\gamma^{\nu}\right]_{+}
=2\eta^{\mu \nu},\;\eta^{\mu \nu}={\rm diag}(1,-1,-1),\;
\mu,\nu=0,1,2$. In this particular case they obey the relations
\begin{equation}
\label{d1a}
\left[\gamma^{\mu}\;,\;\gamma^{\nu}\right]
=-2i\epsilon^{\mu\nu\lambda}\gamma_{\lambda},\qquad
\gamma^{\mu}=\frac{i}{2}\epsilon^{\mu\nu\lambda}
\gamma_{\nu}\gamma_{\lambda}\,.
\end{equation}
In the paper \cite{Git} a path integral 
representation  for the Dirac propagator  was obtained in arbitrary
odd dimensions. In particular, in the case under consideration this
representation reads
\begin{eqnarray}\label{d2}
&&S^{c}=\frac{1}{2}\exp\left(i\gamma^{\mu}
\frac{\partial_{\ell}}{\partial \theta^{\mu}}\right)\int_{0}^{\infty}de_{0}
\int d\chi_{0}\int_{e_{0}} M(e)
De\int_{\chi_{0}}D\chi \int_{x_{in}}^{x_{out}}Dx \int D\pi \int 
D\nu  \\  
&&\times \int_{\psi(0)+\psi(1)=\theta} {\cal D}\psi
\, \exp\left\{i\int_{0}^{1}\left[-\frac{\dot{x}^{2}}
{2e}-\frac{e}{2}m^{2}-g\dot{x}_\mu
A^\mu+iegF_{\mu\nu}\psi^{\mu}\psi^{\nu} 
\right.\right. \nonumber \\
&&\left.\left.
+\chi\left(\frac{2i}{e}\epsilon_{\mu\nu\lambda}\dot{x}^\mu\psi^{\nu}
\psi^{\lambda}-m\right)
-i\psi_{\mu}\dot{\psi}^{\nu}+\pi \dot{e}+\nu \dot{\chi}\right]d\tau
+ \left.\psi_{\mu}(1)\psi^{\mu}(0) \right \}\right|_{\theta=0}\;,\nonumber
\end{eqnarray}
where   $x(\tau)$,  $p(\tau)$,  $e(\tau)$,  $\pi(\tau)$ 
are even and $\psi(\tau),\;\chi_1(\tau),\,\chi_2(\tau),
\;\nu_1(\tau),\nu_2(\tau)$ are odd trajectories, obeying the boundary 
conditions $x(0)=x_{\rm in},\;x(1)=x_{\rm out},\;e(0) =
e_0,\;\chi(0) = \chi_0,\;\psi(0)+\psi(1)=\theta $, and the
notations used are
\[
\chi=\chi_1\chi_2,\;
\nu\dot{\chi}=\nu_1\dot{\chi}_1+\nu_2\dot{\chi}_2,\;
d\chi=d\chi_1\,d\chi_2,\;D\chi=D\chi_1\,D\chi_2,\;D\nu=D\nu_1\,D\nu_2\,.
\]
The measure $M(e)$ is defined by the eq. (\ref{e3}) in the
corresponding dimensions\footnote{We will refer to some formulae from the 
previous sections without specifying that the number of dimensions is
$2+1$ now.}, and
\[
{\cal D}\psi=D\psi\left[\int_{\psi(0)+\psi(1)=0}D\psi\exp
\left\{\int_{0}^{1}\psi_{\mu}\dot{\psi}^{\mu}d\tau\right\}\right]^{-1}.
\]
Integrating over the Grassmann variables in the same way as in the
case of $3+1$  dimensions we get 
\begin{equation}
\label{d3}
S^{c}(x_{out},x_{in})=\frac{i}{2}\int_{0}^{\infty}de_{0}M(e_{0})
\int_{x_{in}}^{x_{out}}Dx\, \Phi[x,e_{0}]\exp\left\{iI[x,e_{0}]
\right\}, 
\end{equation}
where
\begin{eqnarray}
\label{d4}
&&\Phi[x,e_{0}]=\left[\left(m+\frac{i}{e_{0}}\int_{0}^{1}d\tau\,
\epsilon_{\mu\nu\lambda}\dot{x}^{\mu}(\tau){\cal R}^{\nu\lambda}
(g|\tau,\tau)\right)\left(1+\frac{ge_{0}}{4}B_{\alpha\beta}
\gamma^{\alpha}\gamma^{\beta}\right)\right] \nonumber  \\
&&\left.+\frac{i}{2e_{0}}\int_{0}^{1}d\tau\,
\epsilon_{\mu\nu\lambda}\dot{x}^{\mu}(\tau)K^{\nu}_{\quad\alpha}(\tau)
K^{\lambda}_{\quad\beta}(\tau)\gamma^{\alpha}\gamma^{\beta}\right]
\exp\left\{-\frac{e_{0}}{2}\int_{0}^{g}dg^{\prime}\Tr{\cal R}(g)
\star{\cal F}\right\}
\end{eqnarray}
is SF and $I[x,e_{0}],\, {\cal R}(g)\equiv{\cal R}
(g|\tau,\tau^{\prime}),\, K\equiv{K}(\tau),\, B,\, {\cal F}$ are
defined by (\ref{e11}), (\ref{e9a}), and (\ref{e5b}) respectively.

Due to the relations (\ref{d1a}) one can also present SF in the form
\begin{eqnarray}
&&\Phi[x,e_{0}]=\left\{m+\frac{i}{e_{0}}\dot{x}\star{r}(g)
+\left[\left(-i\frac{ge_{0}}{4}m+\frac{g}{4}\dot{x}\star{r}(g)\right)
u_{\alpha}\right.\right. \nonumber \\      
&&+\left.\left.\frac{1}{2e_{0}}\left(\dot{x}\star{T}\right)_{\alpha}\right]
\gamma^{\alpha}\right\} 
\exp\left\{-\frac{e_{0}}{2}\int_{0}^{g}dg^{\prime}\Tr{\cal R}
(g^{\prime})
\star{\cal F}\right\},
\end{eqnarray}
where
\[
r_{\mu}(g)\equiv{r}_{\mu}(g|\tau)=\epsilon_{\mu\nu\lambda}
{\cal R}^{\nu\lambda}(g|\tau,\tau), \qquad
u^{\mu}=\epsilon^{\mu\alpha\beta}B_{\alpha\beta}, \qquad
T_{\mu}^{\quad\rho}=\epsilon_{\mu\nu\lambda}\epsilon^{\rho\alpha\beta}
K^{\nu}_{\quad\alpha}K^{\lambda}_{\quad\beta}.
\] 

\subsection{Dirac propagator in constant uniform field in $2+1$ dimensions}

In the case of constant uniform field $F_{\mu\nu}=const$ one can
calculate the propagator explicitly integrating over the bosonic trajectories.
Following the same way as in Sect.II and taking into account that in 
$2+1$ dimensions 
\[
\frac{i}{2}M(e_{0})\int_{0}^{0}Dx\,\exp\left\{-\frac{i}{2e_{0}}
\dot{x}\star\dot{x}\right\}=\frac{e^{i\frac{\pi}{4}}}{2(2\pi{e}_{0})^
{3/2}}
\]
one gets for the propagator (\ref{d3})
\begin{equation}   
\label{d5}
S^{c}=\frac{e^{i\frac{\pi}{4}}}
{2(4\pi)^{3/2}}\int_{0}^{\infty}
de_{0}\left(\det\frac{\sinh\frac{ge_{0}F}{2}}
{gF}\right)^{-\frac{1}{2}}e^{iI[x_{cl},e_{0}]}\Phi[x_{cl},e_{0}],
\end{equation}
where $x_{cl},\,{\cal R}(g),\,K, B$ are given by 
(\ref{e30}), (\ref{e17}).

The antisymmetric matrices $F_{\mu\nu}$ can be classified by the value
of the invariant $\varphi$ (see Appendix A). In the case
$\varphi^{2}>0$ one can find a Lorentz frame in which the magnetic field
vanishes. On the other hand, $\varphi^{2}<0$ implies that the electric
field vanishes in an appropriate Lorentz frame. The case
$\varphi^2=0,\,F\neq{0}$ corresponds to nonvanishing electric and 
magnetic fields of `equal magnitude' (and this property is Lorentz
invariant). We will consider the case $\varphi^{2}\neq{0}$.
The case $\varphi^{2}=0$ can be easily treated, e.g. taking the 
limit $\varphi\rightarrow{0}$. 

Presenting $\Phi[x_{cl},e_{0}]$ in the case under consideration 
one can avoid the explicit integrations over $\tau$. Indeed, due 
to the specific form of ${\cal R}(g)$ the term containing it in 
(\ref{d4}) vanishes. On the other hand,
\[
\left(\frac{d}{d\tau}-ge_{0}F\right)K=0,
\qquad
\left(\frac{\partial}{\partial\tau}-ge_{0}F\right)\dot{x}_{cl}=0
\]
imply 
\[
\dot{x}_{cl}(\tau)=e^{ge_{0}F(\tau-1)}\dot{x}_{cl}(1), \qquad
K(\tau)=e^{ge_{0}F(\tau-1)}K(1).
\]
Taking into account that $e^{ge_{0}F(\tau-1)}$ is an operator
respecting the scalar product,
one can easily perform the second integration over $\tau$ in
(\ref{d4}). Finally, calculating the determinants involved by 
means of (\ref{a01-13}), one gets
\begin{eqnarray}
&&S^{c}=\sqrt{\frac{i}
{16(2\pi)^{3}}}\int_{0}^{  \infty}
\frac{de_{0}}{\sqrt{e_{0}}}\frac{g\varphi}{\sinh\frac{ge_{0}
\varphi}{2}}e^{iI[x_{cl},e_{0}]}\Phi[x_{cl},e_{0}],\nonumber  \\
\label{d7}
&&\Phi[x_{cl},e_{0}]=\left[m\left(1+\frac{ge_{0}}{4}B_{\alpha\beta}
\gamma^{\alpha}\gamma^{\beta}\right)+\frac{i}{2e_{0}}\epsilon
_{\mu\nu\lambda}\dot{x}^{\mu}(1)K^{\nu}_{\quad\alpha}K^{\lambda}_
{\quad\beta}\gamma^{\alpha}\gamma^{\beta}\right]\cosh\frac
{ge_{0}\varphi}{2}.
\end{eqnarray}

On the other hand one can obtain a representation for the  propagator 
using Schwinger proper-time method (we do not present the calculations
here). Such a representation has the form
\begin{eqnarray}
\label{d7a}  
&&S^{c}(x_{out},x_{in})=\left[\gamma^{\mu}\left(i\frac{\partial}
{\partial{x}^{\mu}_{out}}+gA_{\mu}(x_{out})\right)+m\right]\nonumber \\
&&\times\sqrt{\frac{i}{16(2\pi)^{3}}}\int_{0}^{\infty}
\frac{de_{0}}{\sqrt{e_{0}}}\frac{g\varphi}{\sinh\frac{ge_{0}\varphi}
{2}}e^{iI[c_{cl},e_{0}]}e^{\frac{ge_{0}}{4}F_{\alpha\beta}
\gamma^{\alpha}\gamma^{\beta}}.
\end{eqnarray}
To compare both representations we take the derivative in (\ref{d7a})
and use 
\[
\frac{\partial}{\partial{x}^{\mu}_{out}}I[x_{cl},e_{0}]=
-e_{0}^{-1}\left(\dot{x}_{cl}\right)_{\mu}(1)-gA_{\mu}(x_{out}).
\]
Then one obtains
\begin{equation}
\label{d8}
S^{c}(x_{out},x_{in})=\frac{e^{i\frac{\pi}{4}}}{4(2\pi)^{3/2}}
\int_{0}^{\infty}\frac{de_{0}}{\sqrt{e_{0}}}\frac{g\varphi}
{\sinh\frac{ge_{0}\varphi}{2}}\Psi_{S}(x_{out},x_{in},e_{0})
e^{iI[x_{cl},e_{0}]},
\end{equation}
where
\begin{equation}
\label{d9}
\Psi_{S}(x_{out},x_{in},e_{0})=\left(e_{0}^{-1}\gamma_{\mu}
\dot{x}^{\mu}_{cl}(1)+m\right)e^{\frac{ge_{0}}{4}F_{\alpha\beta}
\gamma^{\alpha}\gamma^{\beta}}.
\end{equation}
Comparing (\ref{d7}) and (\ref{d8})
using the identities (see Appendix B)
\begin{eqnarray}
\label{d10}
&&\exp\left\{\frac{ge_{0}}{4}F_{\alpha\beta}\gamma^{\alpha}\gamma^{\beta}
\right\}=\left(1+\frac{ge_{0}}{4}B_{\alpha\beta}\gamma^{\alpha}
\gamma^{\beta}\right)\cosh\frac{ge_{0}\varphi}{2},\\
\label{d11}
&&\gamma_{\mu}\exp\left\{\frac{ge_{0}}{4}F_{\alpha\beta}\gamma^{\alpha}
\gamma^{\beta}\right\}=\frac{i}{2e_{0}}\epsilon_{\mu\nu\lambda}
K^{\nu}_{\quad\alpha}(1)\,K^{\lambda}_{\quad\beta}(1)
\gamma^{\alpha}\gamma^{\beta}\,\cosh\frac{ge_{0}\varphi}{2}.
\end{eqnarray}
one can verify that both the representations coincide.

\appendix

\section{Some Properties of Antisymmetric Tensors}
\subsection{Antisymmetric tensors in $3+1$ dimensions}
The antisymmetric matrix $F_{\mu\nu}$ formed by the components of the 
stress tensor has in the general case of nonvanishing invariants,
\begin{equation}
\label{eqapp01 01}
{\bf F}=\frac{1}{4}F\sb{\mu\nu}F\sp{\mu\nu}, \qquad
{\bf G}=-\frac{1}{4}F\sb{\mu\nu}F\sp{\ast\mu\nu},
\end{equation}
four isotropic eigenvectors, namely,
\begin{eqnarray}
\label{eqapp01 02}
&F\sb{\mu\nu}n\sp{\nu}=-{\cal E}n\sb{\mu},\quad
&F\sb{\mu\nu}\bar{n}\sp{\nu}={\cal E}\bar{n}\sb{\mu}, \nonumber \\
&F\sb{\mu\nu}m\sp{\nu}=i{\cal H}m\sb{\mu},\quad
&F\sb{\mu\nu}\bar{m}\sp{\nu}=-i{\cal H}\bar{m}\sb{\mu},
\end{eqnarray}
where
\begin{equation}
\label{eqapp01 03}
{\cal E}=\left[\left({\bf F}\sp{2}+{\bf G}\sp{2}\right)\sp{1/2}
-{\bf F}\right]\sp{1/2},\qquad
{\cal H}=\left[\left({\bf F}\sp{2}+{\bf G}\sp{2}\right)\sp{1/2}
+{\bf F}\right]\sp{1/2}.
\end{equation}
The eigenvectors are supposed to be normalized,
\begin{equation}
\label{eqapp01 04}
\bar{n}^{\mu} n_{\mu}=-\bar{m}^{\mu} m_{\mu}=2,
\end{equation}
while all other scalar products vanish,
\begin{equation}
\label{eqapp01 06}
n^{2}=\bar{n}^{2}=m^{2}=\bar{m}^{2}=nm=n\bar{m}=\bar{n}m=\bar{n}\bar{m}=0.
\end{equation}
Then the matrix $F$ can be presented in the form
\begin{equation}
\label{eqapp01 07}
F\sb{\mu\nu}=\frac{{\cal E}}{2}\left(\bar{n}\sb{\mu}n\sb{\nu}-
n\sb{\mu}\bar{n}\sb{\nu}\right)+\frac{i{\cal H}}{2}\left(\bar{m}\sb{\mu}
m\sb{\nu}-m\sb{\mu}\bar{m}\sb{\nu}\right),
\end{equation} 
and, therefore, $F^{2}$ has the spectral decomposition
\begin{equation}
\label{eqapp01 08}
F^{2}={\cal E}^{2}P_{\cal E}+{\cal H}^{2}P_{\cal H}
\end{equation}
where 
\begin{equation}
\label{eqapp01 09}
\left(P_{\cal E}\right)_{\mu\nu}=\frac{1}{2 }\left(\bar{n}_{\mu}
n_{\nu}+n_{\mu}\bar{n}_{\nu}\right),\qquad
\left(P_{\cal H}\right)_{\mu\nu}=\frac{1}{2}\left(\bar{m}_{\mu}
m_{\nu}+m_{\mu}\bar{m}_{\nu}\right)
\end{equation}
are orthogonal projection operators onto some two-dimensional
subspaces,
\begin{eqnarray}
\label{eqapp01 10}
&&P_{\cal E}^{2}=P_{\cal E},\quad P_{\cal H}^{2}=P_{\cal H},\quad
P_{\cal E}P_{\cal H}=P_{\cal H}P_{\cal E}=0,\\
\label{eqapp01 11}
&&P_{\cal E}+P_{\cal H}=\eta,\qquad {\rm tr} P_{\cal E}={\rm tr} 
P_{\cal H}=2.
\end{eqnarray}

\subsection{Antisymmetric tensors in $2+1$ dimensions}

Let $F_{\mu\nu}$ be an antisymmetric matrix in $2+1$ dimensions. 
The antisymmetry implies 
$\tr F=0,\,\det F=0$, so that the sum and the product of the
eigenvalues vanish. The eigenvalues are $0,\,\varphi,\,-\varphi$,
where the real number $\varphi^{2}$ coincides with the invariant
\[
\varphi^{2}=\frac{1}{2}\tr{F}^{2}.
\]
In the case of nonvanishing $\varphi$ there exist three  eigenvectors of
$F$, and $F^{2}$ is proportional to a projection operator $P$ onto
some two-dimensional subspace,
\begin{equation}
\label{a01-12}
F^{2}=\varphi^{2}P,\qquad P^{2}=P,\qquad \tr{P}=2, \qquad
PF=FP=F.
\end{equation} 
Then, for  an even function $h$,
\begin{equation}
\label{a01-13}
h(F)=h(0)\left(1-P\right)+h(\varphi)P,
\end{equation}
while for an odd one 
\begin{equation}
\label{a01-14}
h(F)=\frac{F}{\varphi}h(\varphi).
\end{equation}
The case of vanishing $\varphi$ (and $F\neq{0}$) corresponds to a 
nilpotent matrix, $F^{3}=0$.

\section{Some Identities Involving $\gamma$-Matrices}

\subsection{Gamma-matrix structure of Lorentz transformation 
in the spinor representation}

Let us denote by ${M}(\omega)$ the expression in the RHS of
eq.(\ref{e36}). We are going to
check that the matrix-valued function ${M}(\lambda\omega)$ of a real
parameter $\lambda$ satisfies the differential equation
\begin{equation}\label{a18}
\frac{d}{d\lambda}{M}(\lambda\omega)=-\frac{i}{4}\omega_{\mu\nu}
\sigma^{\mu\nu}{M}(\lambda\omega) 
\end{equation}
and the initial condition
\begin{equation}\label{a19}
{M}(0)=1.
\end{equation}
The latter is trivial, so let us concentrate on the proof of the
equation (\ref{a18}).

With the derivatives
\begin{eqnarray*}
&&\frac{d}{d\lambda}\left(\det\cosh\frac{\lambda\omega}{2}\right)^
{1/2}=\left(\det\cosh\frac{\lambda\omega}{2}\right)\tr\left(\frac{\omega}{4}
\tanh\frac{\lambda\omega}{2}\right),\\
&&\frac{d}{d\lambda}\tanh\frac{\lambda\omega}{2}=\frac{\omega}{2}
\left(\eta-\tanh^{2}\frac{\lambda\omega}{2}\right),
\end{eqnarray*}
one obtains
\begin{eqnarray}\label{a19a}
&&\frac{d}{d\lambda}{M}(\lambda\omega)=\left(\det\cosh\frac{\lambda
\omega}{2} \right)^{1/2}\left\{\left.\frac{1}{4}\tr\left(\omega\tanh
\frac{\lambda\omega}{2}\right)\right.\right.\nonumber \\
&&-\frac{i}{8}\left[\tr\left(\omega\tanh\frac{\lambda\omega}{2}\right)
\tanh\frac{\lambda\omega}{2}+2\omega\left(\eta-\tanh^2\frac{\lambda\omega}
{2}\right)\right]_{\mu\nu}\sigma^{\mu\nu}     \nonumber \\
&&+\frac{1}{8}\epsilon^{\alpha\beta\mu\nu}\left(\tanh\frac{\lambda\omega}{2}
\right)_{\alpha\beta}\left[\frac{1}{4}\tr\left(\omega\tanh\frac{\lambda
\omega}{2}\right)\tanh\frac{\lambda\omega}{2}\right. \nonumber  \\
&&+\left.\left.\omega\left(\eta-\tanh^{2}\frac{\lambda\omega}{2}\right)
\right]_{\mu\nu}\gamma^{5}\right\}.
\end{eqnarray}
Then using the identity
\begin{equation} 
4\epsilon^{\alpha\beta\mu\nu}(ST)_{\alpha\beta}T_{\mu\nu}=
\epsilon^{\alpha\beta\mu\nu}T_{\alpha\beta}T_{\mu\nu}\,\tr{S},
\end{equation}
which is valid for any two matrices\footnote{We remind that, in
accordance with our notation, $\tr
S=S^{\alpha}_{\quad\alpha}=\eta^{\alpha\beta}S_{\beta\alpha}$, etc.}
$S$ and $T$ one can put the derivative (\ref{a19a}) into the form
\begin{eqnarray}\label{a21}
&&\frac{d}{d\lambda}{M}(\lambda\omega)=\left(\det\cosh\frac
{\lambda\omega}{2}
\right)^{1/2}\left\{\left.\frac{1}{4}\tr\left(\omega\tanh\frac{\lambda\omega}
{2}
\right)\right.\right.   \nonumber   \\
&&-\frac{i}{8}\left[\tr\left(\omega\tanh\frac{\lambda\omega}{2}\right)
\tanh\frac{\lambda\omega}{2}+2\omega\left(\eta-\tanh^{2}\frac{\lambda\omega}
{2}\right)\right]_{\mu\nu}\sigma^{\mu\nu}\nonumber              \\
&&+\left.\left.\frac{1}{8}\epsilon^{\alpha\beta\mu\nu}\omega_{\alpha\beta}
\left(\tanh\frac{\lambda\omega}{2}\right)_{\mu\nu}\gamma^{5}\right.\right\}.
\end{eqnarray}
A representation for the product of two $\sigma$-matrices,
\begin{eqnarray}\label{a20}
\sigma^{\alpha\beta}\sigma^{\mu\nu}&&=\eta^{\alpha\mu}\eta^{\beta\nu}
-\eta^{\alpha\nu}\eta^{\beta\mu}    \nonumber \\
&&-i\left(\eta^{\alpha\mu}\sigma^{\beta\nu}+\eta^{\beta\nu}\sigma^{\alpha\mu}
-\eta^{\alpha\nu}\sigma^{\beta\mu}-\eta^{\beta\mu}\sigma^{\alpha\nu}\right)
-\epsilon^{\alpha\beta\mu\nu}\gamma^{5},
\end{eqnarray}
is easily derived from the identities
\begin{eqnarray}
&&\frac{1}{2}\{\sigma^{\alpha\beta},\,\sigma^{\mu\nu}\}=\eta^{\alpha\mu}\eta^
{\beta\nu}-\eta^{\alpha\nu}\eta^{\beta\mu}-\epsilon^{\alpha\beta\mu\nu}\gamma^{5},\\
&&\frac{i}{2}[\sigma^{\alpha\beta},\,\sigma^{\mu\nu}]=\eta^{\alpha\mu}\sigma^
{\beta\nu}+\eta^{\beta\nu}\sigma^{\alpha\mu}-\eta^{\alpha\nu}\sigma^{\beta\mu}
-\eta^{\beta\mu}\sigma_{\alpha\nu}\,.
\end{eqnarray}
Using (\ref{a20}), the well-known identity\footnote{Let us remind
that in our notation
$\gamma^{5}=\gamma^{0}\gamma^{1}\gamma^{2}\gamma^{3}$.}
\begin{equation}
\sigma^{\kappa\rho}\gamma^{5}=\frac{1}{2}\epsilon^{\kappa\rho\tau\sigma}
\sigma_{\tau\sigma},
\end{equation}
and the antisymmetry property $\omega_{\mu\nu}=-\omega_{\nu\mu}$, one
finds
\begin{eqnarray} 
&&-\frac{i}{4}\omega\sigma^{\mu\nu}{M}(\lambda\omega)=\left(\det
\cosh\frac{\lambda\omega}{2}\right)^{1/2}\left[\frac{1}{4}\tr
\left(\omega\tanh\frac{\lambda\omega}{2}\right)\right.\nonumber\\
&&-\frac{i}{64}\epsilon^{\kappa\rho\mu\nu}\left(\tanh\frac{\lambda\omega}
{2}\right)_{\kappa\rho}\left(\tanh\frac{\lambda\omega}{2}\right)
_{\mu\nu}\epsilon_{\alpha\beta\lambda\tau}\omega^{\alpha\beta}\sigma^
{\lambda\tau}                                       \nonumber    \\           
&&-\left.\frac{i}{4}\omega_{\alpha\beta}\sigma^{\alpha\beta}+
\frac{1}{8}\epsilon^{\alpha\beta\mu\nu}\omega_{\alpha\beta}\left(
\tanh\frac{\lambda\omega}{2}\right)_{\mu\nu}\gamma^{5}\right]\,.
\end{eqnarray}
Then, using the identity
\[
\epsilon_{\alpha_{1}\alpha_{2}\alpha_{3}\alpha_{4}}
\epsilon^{\beta_{1}\beta_{2}\beta_{3}\beta_{4}} 
=-\sum_{P}(-1)^{[P]}\delta^{\beta_{1}}_{P(\alpha_{1})}
\delta^{\beta_{2}}_{P(\alpha_{2})}\delta^{\beta_{3}}_{P(\alpha_{3})}
\delta^{\beta_{4}}_{P(\alpha_{4})}\,,
\]
we get
\begin{eqnarray}\label{a22}
&&-\frac{i}{4}\omega_{\mu\nu}\sigma^{\mu\nu}{M}(\lambda\omega)=
\left(\det\cosh\frac{\lambda\omega}{2}\right)^{1/2}\left[\frac
{1}{4}\tr\left(\omega\tanh\frac{\lambda\omega}{2}\right)\right.
                                               \nonumber       \\
&&-\frac{i}{8}\left[\tr\left(\omega\tanh\frac{\lambda\omega}{2}\right)
\tanh\frac{\lambda\omega}{2}+2\omega\left(\eta-\tanh^{2}\frac
{\lambda\omega}{2}\right)\right]_{\mu\nu}\sigma^{\mu\nu}
                                               \nonumber       \\ 
&&+\left.\frac{1}{8}\epsilon^{\alpha\beta\mu\nu}\omega_{\alpha\beta}
\left(\tanh\frac{\lambda\omega}{2}\right)_{\mu\nu}\gamma^{5}\right\}\,.
\end{eqnarray}
The RHS's of (\ref{a21}) and (\ref{a22}) coincide. Therefore ${M}
(\lambda\omega)$ obeys eq.(\ref{a18}). This completes the proof of 
formula (\ref{e36}).

\subsection{Decompositions of some functions on $\gamma$-matrices}

Let us consider the $T$-exponent (\ref{e137})
where $F$ is a uniform constant field, $f$ is a plane-wave
field  and 
(\ref{e115}) takes place. We are going to prove the identities
\begin{eqnarray}
\label{a3-02}
&&\Delta[nx_{cl},e_{0}]=\left(1+\frac{ge_{0}}{4}B_{\alpha\beta}
\gamma^{[\alpha}\gamma^{\beta]}+\frac{g^{2}e_{0}^{2}}{16}
\bar{B}_{\alpha\beta}\bar{B}^{*\alpha\beta}\gamma^{5}\right)
\bar{\Lambda}(e_{0}),       \\
\label{a3-03}
&&\gamma^{\mu}\Delta[nx_{cl},e_{0}]=K^{\mu}_{\quad\nu}(1)
\left(\gamma^{\nu}+\frac{ge_{0}}{4}B_{\alpha\beta}\gamma^{[\nu}
\gamma^{\alpha}\gamma^{\beta]}\right)\bar{\Lambda}(e_{0}),
\end{eqnarray}
where $B$ and $K$ are defined by (\ref{e9a}) for the combination
as it was described while $\bar{B}$, corresponding to the case of 
constant uniform field is given by (\ref{e17}), 
and
\[
\bar{\Lambda}(e_{0})=\exp\left\{-\frac{e_{0}}{2}\int_{0}^{g}dg^{\prime}
\Tr\bar{\cal R}(g^{\prime})\star\bar{\cal F}\right\}=\cosh\frac
{ge_{0}{\cal E}}{2}\cos\frac{ge_{0}{\cal H}}{2}.
\]
Presenting the  $T$-exponent in the form (\ref{e138}) and
using eq.(\ref{e36}) one obtains 
\begin{equation}
\label{a3-05}
\Delta[nx_{cl},e_{0}]=\left(1+\frac{ge_{0}}{4}B_{\alpha\beta}
\gamma^{[\alpha}\gamma^{\beta]}+\frac{g^{2}e_{0}^{2}}{16}\bar{B}_
{\alpha\beta}\bar{B}^{*\alpha\beta}\gamma^{5}\right)\bar{\Lambda}(e_{0})
+Q_{\alpha\beta}\gamma^{[\alpha}\gamma^{\beta]}
\end{equation}
where
\begin{eqnarray}
\label{a3-06}
&&Q=\frac{ge_{0}}{4}\bar{\Lambda}(e_{0})\left(\bar{B}-B\right)+
\frac{1}{4}C,\\
\label{a3-07}
&&C=ge_{0}\int_{0}^{1}d\tau\, e^{\frac{ge_{0}}{2}F(1-2\tau)}
f(nx_{cl}(\tau))e^{-\frac{ge_{0}}{2}F(1-2\tau)}.
\end{eqnarray}
In order to find a convenient representation for $B$ we present $K$, 
which is a solution, obeying (\ref{e111}), to eq.(\ref{e125}, for 
$\phi=nx_{cl}$) in the form
\begin{equation}
\label{a3-08}
K(\tau)=2V(\tau)\left(\eta+V(1)\right)^{-1},
\end{equation}
where
\begin{equation}
\label{a3-09}
V(\tau)=T\,\exp\left\{ge_{0}\int_{0}^{\tau}F^{comb}(nx_{cl}(\tau^{\prime}))
d\tau^{\prime}\right\}.
\end{equation}
is the solution, subjected to (\ref{e111}), to the equation
\begin{equation}
\label{a3-10}
\left[\frac{\partial}{\partial\tau}-ge_{0}F-ge_{0}f(nx_{cl}(\tau))
\right]V(\tau)=0.
\end{equation}
Then, using the defining equation (\ref{e9a}) for $B$ (in which $F$ must 
be understood as $F^{comb}$) and eqs.(\ref{a3-08}), (\ref{a3-09}), one
derives
\begin{equation}
\label{a3-11}
B=\frac{2}{ge_{0}}\left[\eta-2\left(\eta+V(1)\right)^{-1}\right].
\end{equation}
Correspondingly, from eq.(\ref{e17}) we obtain
\begin{equation}
\bar{B}=\frac{2}{ge_{0}}\left[\eta-2\left(\eta+V_{0}(1)\right)^{-1}
\right],\qquad V_{0}(\tau)=e^{ge_{0}F\tau}.
\end{equation}
Solving eq.(\ref{a3-10}) we find
\begin{equation}
\label{a3-12}
V(1)=V_{0}^{1/2}(1)\left(\eta+C+\frac{C^{2}}{2}\right)V_{0}^{1/2}(1),
\end{equation}
by virtue of the nilpotency \cite{b9} of $C$. Then we substitute 
(\ref{a3-12}) into (\ref{a3-11}) and after straightforward
transformations obtain
\begin{equation}
\label{a3-13}
B-\bar{B}=\frac{1}{ge_{0}}\left(\cosh\frac{ge_{0}F}{2}\right)^{-1}
C\left(\cosh\frac{ge_{0}F}{2}\right)^{-1}.
\end{equation}
One can verify, using the transversality (\ref{e105a}) of $a_{\mu}$, 
that\footnote{The projection operators $P_{\cal E}$ and $P_{\cal H}$
are defined in Appendix A.} 
\begin{equation}
\label{a3-14}
C=P_{\cal E}CP_{\cal H}+P_{\cal H}CP_{\cal E}.
\end{equation}
On the other hand, due to the evenness of the function,
\begin{equation}
\label{a3-15}
\left(\cosh\frac{ge_{0}F}{2}\right)^{-1}=
\left(\cosh\frac{ge_{0}{\cal E}}{2}\right)^{-1}P_{\cal E}+
\left(\cos\frac{ge_{0}{\cal H}}{2}\right)^{-1}P_{\cal H}.
\end{equation}
We substitute (\ref{a3-14}) and (\ref{a3-15}) in (\ref{a3-13}) to
get, by virtue of the properties (\ref{eqapp01 10}) of the 
projection operators, 
$P_{\cal E}$ and $P_{\cal H}$,
\begin{equation}
\label{a3-16}
B-\bar{B}=\frac{1}{ge_{0}}\left(\cosh\frac{ge_{0}{\cal E}}{2}\,
\cos\frac{ge_{0}{\cal H}}{2}\right)^{-1}C.
\end{equation}
Finally, inserting  (\ref{a3-16}) into (\ref{a3-06}) and using
\begin{equation}
\label{a3-17}
\bar{\Lambda}(e_{0})=\left(\cosh\frac{ge_{0}{\cal E}}{2}\,
\cos\frac{ge_{0}{\cal H}}{2}\right)^{-1},
\end{equation}
one finds that $Q=0$ and the identity (\ref{a3-02}) takes place.

Going to the identity (\ref{a3-03}) we use (\ref{a3-02}), (\ref{e9b}),
the identity 
\[
\gamma^{\mu}\gamma^5=-\frac{1}{3!}\epsilon^{\mu}_{\quad\kappa\rho\sigma}
\gamma^{\kappa}\gamma^{\rho}\gamma^{\sigma}
\]
and the antisymmetry $B_{\alpha\beta}=B_{[\alpha\beta]}$ to bring the
LHS into the form
\begin{eqnarray}
\label{a3-18}
&&\gamma^{\mu}\Delta[nx_{cl},e_{0}]=\left[\left(\eta^{\mu}_{\quad\beta}+
\frac{ge_{0}}{2}B^{\mu}_{\quad\beta}\right)\gamma^{\beta} 
\right.\nonumber\\
&&\left.
+\frac{ge_{0}}{4}\left(\eta^{\mu}_{\quad\nu}B_{\alpha\beta}-\frac
{ge_{0}}{4!}B_{\rho\sigma}B^{*\rho\sigma}\epsilon^{\mu}_{\quad\nu\alpha
\beta}\right)\gamma^{[\nu}\gamma^{\alpha}\gamma^{\beta]}\right]
\bar{\Lambda}(e_{0}).
\end{eqnarray}
>From (\ref{a3-08}) and (\ref{a3-11}) one obtains
\begin{eqnarray}
\label{a3-19}
&&K(1)=\eta+\frac{ge_{0}}{2}B,       \\
\label{a3-20}
&&K_{\mu[\nu}(1)B_{\alpha\beta]}
=\eta_{\mu[\nu}B_{\alpha\beta]}+\frac{ge_{0}}{2}B_{\mu[\lambda}
B_{\alpha\beta]}. 
\end{eqnarray}
Due to the antisymmetry of $B$,
\[
B_{\mu[\lambda}B_{\alpha\beta]}=B_{[\mu\nu}B_{\alpha\beta]}=
-\frac{2}{4!}B_{\kappa\rho}B^{*\kappa\rho}\epsilon_{\mu\nu
\alpha\beta},
\]
and, substituting in (\ref{a3-20}), we get
\begin{equation}
\label{a3-21}
K_{\mu[\nu}(1)B_{\alpha\beta]}=\eta_{\mu[\nu}B_{\alpha\beta]}
-\frac{ge_{0}}{4!}B_{\kappa\rho}B^{*\kappa\rho}\epsilon_
{\mu\nu\alpha\beta}.
\end{equation}
Finally, we use (\ref{a3-19}), (\ref{a3-21}) in (\ref{a3-18}) to get
(\ref{a3-03}).

\subsection{Identities involving $\gamma$-matrices in $2+1$
dimensions}

To prove the identity (\ref{d10}) let us introduce
\[
z^{\mu}=\epsilon^{\mu\nu\lambda}F_{\nu\lambda}, \qquad
z^{2}=-4\phi^{2}
\]
and transform the LHS of (\ref{d10}) using (\ref{d1a}),
\[
\exp\left\{\frac{ge_{0}}{4}F_{\mu\nu}\gamma^{\mu}\gamma^{\nu}
\right\}=\cosh\frac{ge_{0}\varphi}{2}\left(1-\frac{iz\gamma}{2\varphi}
\tanh\frac{ge_{0}\varphi}{2}\right).
\]
Taking into account eqs.(\ref{e9a}), (\ref{a01-14}) and the relation 
$iz\gamma=-\gamma{F}\gamma$, one gets (\ref{d10}).

Multiplying (\ref{d11}) by $K(1)$ and using (\ref{d1a}) one transforms 
(\ref{d11}) into the equivalent identity
\begin{equation}
\label{a3-30}
K^{\mu}_{\quad\lambda}(1)\gamma^{\lambda}e^{\frac{ge_{0}}{4}F_{\alpha\beta}
\gamma^{\alpha}\gamma^{\beta}}=\gamma^{\mu}\cosh{\frac{ge_{0}\varphi}{2}}   
\det{K}(1),
\end{equation}
which we are going to prove. Taking into account the identity
\[
\gamma^{\lambda}e^{\frac{ge_{0}}{4}F_{\alpha\beta}\gamma^{\alpha}
\gamma^{\beta}}=\left(e^{-\frac{ge_{0}}{2}F}\right)^{\lambda}_
{\quad\rho}\gamma^{\rho},
\]
which can be easily derived from (\ref{d10}) and using (\ref{e9a})
one transforms the LHS of (\ref{a3-30}),
\begin{equation}
\label{a3-31}
K^{\mu}_{\quad\lambda}(1)\gamma^{\lambda}e^{\frac{ge_{0}}{4}
F_{\alpha\beta}\gamma^{\alpha}\gamma^{\beta}}=\left(\cosh
\frac{ge_{0}}{2}\varphi\right)^{-1}\gamma^{\mu},
\end{equation}
Calculating the determinant
\[
\det{K}(1)=\left(\cosh\frac{ge_{0}\varphi}{2}\right)^{-2},
\]
one finds that the RHS of (\ref{a3-31}) coincides with that of 
(\ref{a3-30}).    

{\large{\bf Acknowledgments}}

D. M. Gitman thanks Brazilian Foundation CNPq for support.
S.~I.~Zlatev thanks the Department of Mathematical Physics of the
University of S\~ao Paulo for hospitality.

\end{document}